\documentstyle[12pt,aaspp4]{article}
\everymath{\rm}
\everydisplay{\rm}
\received{}
\revised{}
\accepted{}
\cpright{}{}\journalid{}{}
\articleid{}{}
\paperid{}
\slugcomment{Invited Review, {\it Publications of the Astronomical Society of the Pacific}}
\lefthead{Henry \& Worthey}
\righthead{Abundance Profiles}
\begin{document}
\title{The Distribution Of Heavy Elements In Spiral And Elliptical Galaxies}

\author{R.B.C. Henry}
\affil{Department of Physics \& Astronomy, University of
Oklahoma, Norman, OK  73019; henry@mail.nhn.ou.edu}

\and

\author{Guy Worthey}
\affil{Department of Physics \& Astronomy, St. Ambrose University, Davenport, IA, 52803; gworthey@saunix.sau.edu}

\begin{abstract}

In large disk and spheroidal galaxies spatially resolved abundance
information can be extracted by analysis of either emission lines,
absorption lines, or both, depending on the situation. This review
recaps significant
results as they apply to non-dwarf galaxies, including the Milky
Way, spiral disks and bulges, and elliptical and lenticular galaxies.
Methods for determining abundances are explained in appendices. 

Conclusions that span the galaxy types treated here are as follows. All
galaxies, on average, have heavy element abundances (metallicities)
that systematically decrease outward from their galactic centers while
their global metallicities increase with galaxy mass. Abundance
gradients are steepest in normal spirals and are seen to be
progressively flatter going in order from barred spirals, lenticulars,
and ellipticals. The
distribution of abundances $N(Z)$ vs. $Z$ is strongly peaked compared
to simple closed-box model predictions of chemical enrichment in all
galaxy types. That is, a ``G dwarf problem'', commonly known in the
solar cylinder, exists for all large galaxies.

For spiral galaxies, local metallicity appears to be correlated with
total (disk plus bulge) surface density. Examination of N/O versus O/H
in spiral disks indicates that production of N is dominated by primary
processes at low metallicity and secondary processes at high
metallicity. Carbon production increases with increasing
metallicity. Abundance ratios Ne/O, S/O, and Ar/O appear to be
universally constant and independent of metallicity, which argues
either that the IMF is universally  constant or that these ratios are not sensitive
to IMF variations. In the Milky Way, there is a rough age-metallicity
trend with much scatter, in the sense that older stars are more
metal-poor.

In elliptical galaxies, nuclear abundances are in the range
[Z/H] = 0.0 to 0.4, but the element mixture is not scaled-solar. In
large elliptical galaxies [Mg/Fe] is in the range 0.3 to 0.5,
decreasing to $\approx$0 in smaller elliptical galaxies. Other light
elements track the Mg enhancement, but the heavier Ca tracks Fe. Velocity
dispersion appears to be a key parameter in the modulation of [Mg/Fe],
but the cause of the connection is unclear.

\end{abstract}


\section{Introduction}

Galactic chemical evolution is the proportional buildup
of helium and heavy elements or metals, i.e. elements other than hydrogen
and helium, within a galaxy over time as a result of the
continuous manufacture and expulsion of these elements by
resident stars. The topic
concerns itself not only with global or pan-galactic changes but
also with regional ones within spatially-resolved galaxies.

The essence of chemical
evolution can be illustrated by imagining a closed box containing
interstellar gas of primordial composition situated at an
arbitrary location within a galaxy. As portions of the gas collapse, fusion
processes within the stars that are formed convert hydrogen into
heavier elements, and this chemically enriched material is
subsequently expelled into the interstellar medium through
stellar winds, planetary nebula formation, or supernova
eruptions. As this cast-off matter mixes with the surrounding
medium, the composition of the latter changes such that the
abundances of helium and the heavy elements increase relative to hydrogen.
As a second generation of stars forms from this
enriched material, the new stars possess a greater fraction of
heavy elements than their predecessors. Thus, the enrichment cycle
continues until enough material has been locked up in
stellar remnants that the star formation process, which depends
upon the availability of interstellar gas, is finally damped.
A galaxy, then, can be thought of as an ensemble of these boxes.

The real picture is more complicated, of course. The boxes have
no walls, and as such are open to matter exchange with their
surroundings in all directions. 
Nevertheless, the simple model does suggest some of the crucial processes
which must be understood if we are to have a comprehensive
understanding. For example, we must know the details of star
formation and evolution, stellar nucleosynthesis and the rate of
heavy element production, the details of stellar death and matter
ejection, and the efficiency with which ejected material is mixed
into the interstellar medium.

Studies of galactic chemical evolution involve an interplay
between 1)~global and/or spatially resolved abundances, sometimes
as a function of time, for one or more galaxies; and 2)~models based
upon a set of input parameters determined by the physics being
tested. Observed abundances provide two basic sorts of
information. First,
ratios of heavy elements relative to hydrogen, such
as O/H or Fe/H,
serve as gauges of how far chemical evolution has progressed in a
system, because they
measure the extent to which hydrogen has been converted to
heavier elements. As such, these ratios are particularly
sensitive to the rate at which gas is cycled through stars, i.e. the
star formation rate, and how that rate may have changed with
time.  Second,
ratios of two heavy elements, such as N/O or O/Fe, provide information
about differential elemental production by stars. That is, at
what rate, say, is nitrogen produced relative to oxygen, or
oxygen relative to iron? The answer here is tied to the
production rates of individual elements as a function of stellar
mass weighted by the relative number of stars at each mass (the initial mass function; IMF) as
well as to the history of star formation.
Also, there is an element of time involved in all of
this. For example, abundances measured in a star
reveal enrichment levels at
the time the star formed. In summary, chemical evolution can be
traced indirectly by associating abundance patterns within a galaxy with local conditions, where the latter ultimately depend on time, and directly
by observing abundances in stars of different ages or in galaxies of
different look-back times.

The primary goal of this review is to describe the state of affairs
concerning observed abundance patterns in
galaxies. Because of author
expertise, emphasis is placed on abundance patterns in spiral
disks and elliptical galaxies as derived from emission line
analyses and photometric
indices, respectively. However, for completeness and continuity,
we also describe and compare the complimentary results
provided by stellar abundance work in the Milky Way and nearby
galaxies.
Our elemental scope is confined to those elements between carbon
and iron on the periodic table (6$\le$Z$\le$26),
i.e. those elements which are the
most readily observed and for which there is the most
information.
Discussions of helium and the light elements are better
taken up in the context of Big Bang nucleosynthesis, and for
this the reader is urged to consult Chapter~4 of Pagel (1997)
and references therein for recent discussions of this topic.
Likewise, elements beyond iron have been studied in part by
Edvardsson et al. (1993), Wheeler, Sneden, \& Truran (1989), and
McWilliam (1997).

Numerous reviews of galactic chemical evolution
and abundance patterns are available in the literature.
An excellent, approachable introduction to the subject of chemical
evolution is given in the comprehensive review by Tinsley (1980).
The textbook by Pagel (1997) treats numerous topics
related to galactic chemical evolution
and the synthesis of elements. Additional
material on observations and abundance studies in galaxies can
be found in several recent conference proceedings, in particular
Friedli et al. (1998) and Walsh \& Rosa (1999).  Other useful works specifically
treating element synthesis include books by Clayton (1983), Rolfs \&
Rodney (1988) and Cowley (1995), the review by Trimble (1991),
and the conference
proceedings by Edmunds \& Terlevich (1992) and Prantzos,
Vangioni-Flam, \& Cass{\'e} (1993). Finally, QSO absorption line systems are enabling chemical evolution studies to be carried out through the study of abundances as a function of look-back time. While these systems are beyond our scope, interested readers are urged to consult Lauroesch et al. (1996), Lu et al. (1996), and Pettini et al. (1999).

We begin with a discussion of abundances derived from emission lines
in spiral galaxies, including the Milky Way, in {\S}2. In {\S}3 we
turn to stars both inside and outside of the Milky way, while abundances in
elliptical galaxies from photometric integrated light are treated in
{\S}4. A summary is given in {\S}5. Appendices explain techniques used
to derive abundances from emission lines (Appendix A), stellar absorption
lines (Appendix B), and the integrated starlight of composite systems
(Appendix C). Unless otherwise stated, elemental abundances and ratios referred to in this review are by number, not mass.

\section{Abundance Patterns In Spiral Galaxies From Emission-Line Objects}

Sampling a galaxy's interstellar medium directly provides a snapshot of
the {\it current} abundance picture at the location being tested, in contrast to stellar abundances which for the most part are indicative of interstellar abundances at the time that the star formed.
The most straightforward way of determining interstellar
abundances is through the analysis of emission
spectra produced by gas heated by nearby hot stars with
continua rich in photons having wavelengths shortward of
912{\AA}, i.e. the ionization edge of hydrogen. Such stars have
effective temperatures exceeding 30,000K and spectroscopically
belong to the O and early B classes.  Object types with these
conditions include H~II regions and planetary nebulae, reviews
of which can be found
in Shields (1990), Vila-Costas \& Edmunds (1992),  and Zaritsky, 
Kennicutt, \& Huchra (1994) for H~II regions and Peimbert (1990), 
Henry (1990), Perinotto (1991), Clegg (1993), and 
Habing \& Lamers (1997) for planetary nebulae.
Old supernova remnants in which the stellar ejecta have
completely mixed with the interstellar medium in principle
represent a third type of emission line probe, since in this
case the interstellar gas is heated by the shockwave producing
emission lines.
Often, however, full abundance studies are precluded by limited
spectral coverage even within the optical (W.P. Blair, private communication),
and thus there are far fewer
abundance results available.

Ionized gases of the types just mentioned maintain temperature
equilibrium in most cases by radiating photons at discrete wavelengths
following
recombination or collisional excitation processes involving
ion-electron encounters. Spectra of
these objects can then be analyzed to provide abundance,
temperature, and density information.
Measured from the ground in most cases, the strengths
of the resulting emission lines can be converted to
ionic and elemental abundances of He, C, N, O, Ne, S, and Ar
using techniques described in Appendix~A, which includes a table
listing a number of prominent emission features.
While the resulting
abundances refer to levels in the gas phase only, Savage \&
Sembach (1996) indicate that none of these elements is
expected to be heavily partitioned into the solid phase in the
form of dust. Thus, gas phase abundances should represent total
values reasonably well.

Due to their size and therefore their accessibility in external
galaxies, most of the abundance data from emission-line systems
useful in chemical evolution studies relate to H~II
regions, which because of their association with recent star
formation are located in spiral
disks and irregular galaxies. This section focuses on
abundance patterns in spirals.

It should be noted that abundances discussed are
taken directly from the sources listed; no
attempt has been made to homogenize them by recalculating the
abundances in a consistent way. In general, differences
in techniques and atomic data employed produce ranges in
abundances which are smaller than observational uncertainties in
the line strengths, and we believe that presenting unhomogenized
data still provides a realistic representation of patterns and
an opportunity to see the big picture.

\subsection{Metallicity Gradients In Spiral Disks}

Metallicity is the fraction by mass of all elements heavier than He in a
system and is the primary indicator of chemical
evolution as stars convert H into heavier elements and seed their
environments with the products. 
Oxygen is the metallicity
tracer of choice in the interstellar medium. Cosmically, its
relative abundance surpasses all elements but H and He. Its
relatively small depletion (Snow \& Witt 1996; Savage \& Sembach
1996) means it is
present almost entirely in the gas phase.
Hot, ionized gas in the vicinity of hot stars or energetic shock
waves give rise to H~II regions, planetary nebulae, or supernova
remnants whose spectra usually display prominent emission lines
of oxygen. This contrasts sharply with the situation for old
stars, for example, where absorption features of iron (heavily
depleted onto grains in the interstellar medium) are prominent in stellar
spectra due to the presence of optimal temperatures. Thus, iron
is usually employed as a metallicity indicator when old stars are
the probes.

Differences in the appearances of H~II region spectra as a function of 
galactocentric distance were first noticed by Aller (1942) in his study 
of M33. Thirty years later, Searle (1971) connected similar differences 
across disks of several Sc galaxies with systematic changes in heavy element 
abundances. Early abundance gradient work in spirals is reviewed by 
Pagel \& Edmunds (1981), while the more contemporary picture is
available in Friedli et al. (1998).

\subsubsection{Metallicity Gradients In The Milky Way Galaxy}

The disk
of the Milky Way Galaxy (MWG) is arguably the most active and rapidly evolving
region of our galaxy in the chemical sense. Probing it, though,
is complicated by the presence of dust along all lines of sight
within the disk, preventing radiation, particularly in the
ultraviolet, from readily penetrating it. From our location within
the disk, observing emission line objects is restricted not only
from effects of reddening, but the presence of dust limits the
distance over which we can probe. Despite these restrictions,
large amounts of data are now available for H~II regions and planetary
nebulae within the disk of the MWG. The distance range from the
sun in both the directions of the galactic center and anticenter
has been extended by observations in the infrared, where
extinction is at a minimum.
 
Table 1 summarizes the data for emission line objects
in the disk of the MWG and compiled here.  The columns in
order indicate the type of object and spectral region (optical,
far infrared, or radio) followed by the name of the
first author on the paper for the data source, the total number of and
galactocentric distance range for objects included in each study,
and finally an indication of the number of data points for each
abundance ratio which were available in each study. 

Six of the eight studies in Table~1 are based upon H~II regions.
The seminal study by Shaver et al. (1983) explored a range in
galactocentric distance centered on the sun. More recent
studies by Fich \& Silkey (1991), V{\'i}lchez \&
Esteban (1996), and Rudolph et al. (1997)
focused on the anti-center direction, while Simpson
et al. (1995) and Afflerbach et al. (1997) studied objects
toward the center of the Galaxy. Although the optical studies were
all ground-based, the far infrared work in all three cases was
carried out using the Cryogenic Grating Spectrometer aboard the
Kuiper Airborne Observatory to observe emission lines such as
[N~III] 57$\mu$m, [O~III] 52,88 $\mu$m, and [S~III] 19,33 $\mu$m.

The remaining two papers in Table~1 were based on studies of
planetary nebulae (Maciel \& K{\"o}ppen 1994) and
supernova remnants (Fesen, Blair, \& Kirshner 1985). Planetary
nebulae comprise ejected material from evolved intermediate mass
stars, and thus generally have an abundance profile which differs somewhat from
that of the progenitor star at the time of birth due to contamination of the nebula by products of stellar nucleosynthesis. However, the contamination affects primarily
helium, carbon, and nitrogen, and thus abundances of oxygen,
sulfur, and argon in the nebula are expected to be good measures
of the levels of those elements in the nascent progenitor star. In particular,
Type~II planetary nebulae (Peimbert 1978) have progenitors of
small enough mass that oxygen is not expected to have been
altered by CNO processing, yet they are disk objects, based upon
their kinematics. Finally, the study of old supernova remnants by Fesen et
al. measured abundances in disk objects over a radial range
similar to that of Shaver et al.

Figure~1 plots
O/H versus galactocentric distance in kiloparcsecs by author,
where 12+log(O/H) is used to represent oxygen\footnote{H~II region distances in Shaver et al.'s data
have been recomputed using their reported radial velocities and
longitudes, eq.~9.3 in Binney \& Merrifield (1998), and
assumptions that $R_{\sun}=8.5$kpc and the average circular
velocity is 240~km/s across the relevant portion of the disk.}.
Representative
error bars are shown in the lower left panel.
Solid lines show first order least squares
fits for which the fitting parameters are
given in Table~2, with the gradient G expressed
in dex/kpc, the absolute abundance A$_{8.5}$ given as 12+log(O/H) at the
solar circle (8.5~kpc), and c is the correlation coefficient. Dot-dashed lines show the composite fit from Table~2 for reference.
Note that reported upper and lower limits on abundance ratios were
generally not used in our compilation. We also point out that the larger
scatter in the Simpson et al. data is probably due in large part
to inferring O/H from O$^{+2}$/S$^{+2}$ observations plus an
assumption of a constant S/O ratio, while the scatter in the Fesen et
al. results for supernova remnants may be due to their using abundance-line strength diagnostic diagrams taken from shock models in the literature for estimating abundances. 

Combining several data sets allows a visual comparison to be
made among them as well as a test of the robustness of the trend
exhibited by a single set. Figure 2 shows
12+log(O/H) versus galactocentric distance in kpc for the nebular
data sets of Shaver et al., Afflerbach et al., Maciel \&
K{\"o}ppen, V{\'i}lchez \& Esteban, Fich \& Silkey,
Rudolph et al., and Fesen et al. We also present the B
star results from Smartt \& Rolleston (1997; filled circles) and Gummersbach et al. (1998; open circles).  The
sun's position (Grevesse \& Noels 1993) is indicated with an `x',
while the error bars in the lower left show typical
observational uncertainties for all of the data.  A monotonic
decrease in oxygen abundance with galactocentric distance is
clearly present in the Milky Way disk. A simple least squares fit to
all points except the B~stars indicates a slope of
-0.06($\pm$0.01), A$_{8.5}$ of 8.68 ($\pm$0.05), and a
correlation coefficient of -0.63. The data from Afflerbach et al.
and V{\'i}lchez \& Esteban extend the trend of the main body of
data toward the galactic center and anti-center, respectively.
The uncertainty of $\pm$0.2 dex in oxygen abundance is consistent
with observational uncertainty, and thus there is no indication
of real abundance scatter at a constant radial distance, in
accord with findings of Kennicutt \& Garnett (1996) in their
study of M101.

Several additional points are illustrated in Fig.~2. First, notice that
the B star oxygen abundance trend is not noticeably different from the one defined by nebular
data; in fact their gradients are very similar to the nebular results.
This represents a major development, as previous attempts to infer the
disk O/H distribution from B stars (Gehren et al. 1985; Fitzsimmons et
al.  1992; Kilian-Montenbruck et al. 1994; Kaufer et al. 1994)
indicated the absence of a gradient. Smartt \& Rolleston speculate that
sample size was the culprit in obscuring the gradient in most of the
previous studies.

Next, the oxygen abundance distribution implied by planetary nebulae is
indistinguishable from the one from H~II regions. This would seem to
confirm the value of PNe to trace disk metallicity and at the same time
perhaps reduce the concern about diffusion (Wielen et al. 1996), i.e.
that positions of PN progenitors shift radially during their lives, and
thus PN abundances do not represent ISM conditions at their present
galactocentric distances.

An interesting wrinkle in the MWG metallicity
gradient picture is the possibility that the gradient flattens
beyond 10~kpc. Results for the three anti-center studies are presented in a
single panel in Fig.~1. V{\'i}lchez \& Esteban conclude that their data
support the presence of a flattened oxygen gradient in the outer
galaxy, although results from Fich \& Silkey and Rudolph et al.
neither support nor counter this claim. Recently,
Maciel \& Quireza (1999) have expanded and updated their sample
to include PNe with larger galactocentric distances than those
presented here. Like V{\'i}lchez \& Esteban above, they find
evidence for a gradient which flattens beyond 12~kpc. A
flattened gradient is both a controversial and interesting conjecture
and is tied to the
dynamics and mass distributions in the disk (Zaritsky 1992;
Moll{\'a} et al. 1996; Samland, Hensler, \& Theis 1997), and we
briefly return to this point in {\S}2.4 in our general discussion of
spiral abundance gradients.

Finally, we have plotted predictions of four chemical evolution models
of the present-day disk for comparison with the data. The dashed line
shows an analytical result based on the ``simple model'' from Pagel
(1997; eq.~8.14), while detailed numerical model results are shown from
Samland, Hensler, \& Theis (1997; solid line), Ferrini et al. (1994;
dot-dashed line), and K{\"o}ppen (1994; long-dashed line), where
K{\"o}ppen (private communication) employed a quadratic star formation
law but recalculated his model for a radial flow velocity of
0.3km/s\footnote{The unpublished K{\"o}ppen model also assumes a disk age of
15~Gyr along with exponentially decreasing infall both in time (5~Gyr
scale) and galactocentric distance (4 kpc scale)}.  All models are
scaled so as to match our composite interstellar oxygen abundance of
8.68 at the solar circle (see Table~2).  Note that the Samland et al.
model predicts a gradient flattening outward from around 11~kpc, the
result of (according to them) mass loss of long-living metal-poor
intermediate mass stars and additional infall of low metallicity gas in
equilibrium with metal enrichment from condensation of intercloud
medium.

\subsubsection{Metallicity Gradients In External Galaxies}

Results from numerous surveys of spiral galaxy abundance patterns
show clearly
that most spiral disks possess negative gradients qualitatively
similar to the one in the Milky Way. Large surveys of O/H in
extragalactic H~II regions include those of Mc~Call (1982; 40
galaxies; see also Mc~Call, Rybski, \& Shields 1985), Vila-Costas \& Edmunds (1992; 32 galaxies), and
Zaritsky, Kennicutt, \& Huchra (1994; 39 galaxies). We can add
to those the recent
studies by Ferguson, Gallagher, \& Wyse (1998) and
van~Zee et al. (1998), both of which explored the outer
regions of spirals, where star formation rates are much lower and
the regions are less advanced chemically. Vila-Costas \& Edmunds
reprocess reduced line strengths from the literature
to obtain their abundances, while the other authors use
primarily their own data for their studies. All these studies
are based on optical spectra.

As an example of abundance patterns in two external spirals, in Fig.~3
we present a comparison of results for NGC~628 and M33 from Zaritsky
et al. (1994) with Milky Way data from Shaver et al. (1983),
Afflerbach et al. (1997), and V{\'i}lchez \& Esteban (1996), where
12+log(O/H) is plotted against galactocentric distance. The latter
quantity has been normalized to the respective galaxy's isophotal
radius R$_o$\footnote{The isophotal radius R$_o$ is the radial distance from
the galactic nucleus at which the declining disk surface brightness
reaches 25 mag/arcsec$^2$. Comparisons of data among galaxies are made
most frequently using the isophotal radius, but one could also use the
effective radius, i.e. the radius of an aperture admitting one-half of the
light from the disk, or kiloparsecs. To add to the confusion,
literature sources for (non-nebular) bulge or elliptical galaxy
abundances usually express gradients as
$\Delta$log~$Z$/$\Delta$log~$R$! This last notation gives nonsensical
abundances at the nucleus of a galaxy, but seems to match observed
profiles out to the observational limit, which is usually far short of the isophotal radius for integrated starlight spectroscopy.} to account for size
variations among galaxies.  R$_o$ for the Milky Way disk was taken
from de~Vaucouleurs \& Pence (1978).  Gradient slopes determined from
least squares fits are given in the figure legend.  The two external
galaxies clearly resemble the Milky Way in possessing negative
abundance grandients.

A much larger collection of abundance plots for individual spirals can be
found in Zaritsky et al. (1994).
We have extracted results from that paper
and plotted them in Figure~4a, where characteristic
abundances\footnote{The characteristic abundance is the
abundance at 0.4R$_o$ as determined by a least squares fit to the
data. See Zaritsky et al. (1994).}
(top panels) and gradient slopes in dex/R$_o$ (bottom panels)
are shown as functions of galaxy
morphological type (T type), absolute blue magnitude M$_B$, and
maximum circular velocity V$_c$ in km/s.  All three of these
parameters track galaxy mass, where
smaller T type indices, more luminous integrated blue magnitudes,
and larger rotational velocities generally correspond with more
massive spirals. Normal
(SA) and barred (SB) spirals are shown separately using filled and
open symbols, respectively. Abundance parameters for the
Milky Way composite fit from Table~2 are indicated in Fig.~4a  with plusses,
where we have adopted T=4, M$_B$=-20.08, and R$_o$=11.5kpc
(de~Vaucouleurs \& Pence 1978), along with V$_c$=220~km/s
(Kochanek 1996).

Two important points are implied by Fig.~4a: (1)~Characteristic
abundances increase with galaxy mass, while gradient slopes are
uncorrelated with this parameter; and (2)~Characteristic
abundances in normal and barred spirals are indistinguishable, but
barred spirals appear to have flatter (less negative) gradients. 
Both of these results have been noted previously.  Garnett \&
Shields (1987) plotted characteristic O/H values against galaxy
mass for numerous spirals and found a direct correlation between
these two parameters, while Pagel et al. (1979) first suggested
that barred spirals may have flatter gradients, a pattern clearly
borne out in the more extensive work by Martin \& Roy (1994),
who
relate gradient slope to bar strength, a quantity which measures
bar ellipticity.  Martin \& Roy find direct relations between the slope of
the oxygen abundance gradient of a barred spiral and the galaxy's
bar strength (ellipticity) and length in the sense that stronger bars are accompanied by flatter gradients.  This empirical result is
consistent with radial flow models of chemical evolution in which
the presence of a bar enhances large-scale mixing over the
galaxy's disk, damping radial abundance variations.

Interestingly, if gradient slope in dex/kpc (as opposed to
dex/R$_o$ shown here) is plotted versus M$_B$ (see Garnett 1998)
a correlation appears such that more luminous galaxies have
flatter slopes. The dependence of slope behavior on normalization
is no doubt related to the fact that R$_o$ for luminous galaxies
tends to be longer in kiloparsecs. Since the vertical scatter in
the lower panel of Fig.~4a is comparable to observational
uncertainties, this may imply a universal gradient in dex/R$_o$, which
in turn could be associated with similar timescales for viscous
angular momentum transport and star formation, producing
exponential gradients in surface density and abundances (Lin \& Pringle 1987; Yoshii \& Sommer-Larsen 1989).

We illustrate explicitly the correlation between galaxy mass and
characteristic abundance
in Fig.~4b, where we plot 12+log(O/H) at one effective radius versus the
log of the galaxy mass in solar units for a sample of spiral
galaxies. Abundance data are from Garnett \& Shields (1987),
Skillman et al. (1996), and Henry et al. (1996). Sources for
galaxy masses, which for the most part are inferred from rotation curves, are given in Henry et al. The two points connected by a horizontal
line are for NGC~753 whose mass was determined for H$_o$ values
of 50 and 100 km/s. The least squares fit to the data, shown with a solid line, indicates that $12+log(O/H)=3.79+0.47 \times logM$, where $M$ is in solar masses. Note that this relation ignores the low surface brightness spirals (McGaugh 1994) which appear to have low metallicity but high mass. These objects are discussed briefly in {\S}2.1.3.

Finally, Figure~5 shows the observed relation between 12+log(O/H) and total (disk + bulge) surface density, $\Sigma$, in M$_{\sun}$/pc$^2$ at the corresponding location from
Vila-Costas \& Edmunds (1992; their Fig.~7d), where open and
closed squares represent H~II regions residing in late (Scd-Irr)
and early (Sab-Sc) spirals, respectively.  Vila-Costas \& Edmunds assumed that the mass distribution follows the light distribution and employed a constant mass-to-light ratio for each galaxy, the latter determined from a rotation curve (see Vila-Costas \& Edmunds and Mc~Call 1982 for details).
The scatter is
consistent with observational uncertainty, and thus we see a
clear positive correlation between abundance and local surface density
in spirals, with earlier spirals generally possessing higher abundances per unit
surface density.

\subsubsection{Assorted Issues About Galaxy Metallicity}

Other issues concerning abundance
gradients include questions about gradients
perpendicular to the disk as well as azimuthal distributions,  abundance patterns in low
surface brightness galaxies, effects of cluster environment on
gradients, the mathematical form of abundance profiles, and results of extragalactic planetary nebula studies. We
treat these topics briefly.

{\it A negative vertical gradient in O/H in the Milky Way} is
suggested by planetary nebula studies.  Abundance data compiled by
Kaler (1980) for PNe ranging in height above the disk from less
than 0.4~kpc to greater than 1~kpc show a decrease in O/H with
increasing height above the plane.  A comparison of more recent
studies of PNe close to the plane (Perinotto 1991), greater than
300pc above the plane (Cuisinier et al. 1996), and in the halo
(Howard, Henry, \& McCartney 1997) shows averages of 12+log(O/H) for these
three samples of 8.68, 8.52, and 8.02 respectively, qualitatively
consistent with Kaler.

Thorough tests for {\it azimuthal gradients in spiral disks} have
yet to be carried out.  One example of apparent O/H asymmetry is
discussed by Kennicutt \& Garnett (1996) in their study of M101. 
They find that H~II regions located along a spiral arm southeast
of the major axis have a lower oxygen abundance by 0.2-0.4~dex
compared with H~II regions on the opposite side.

{\it Global metallicities in low surface brightness galaxies} are
generally found to be subsolar by roughly a factor of three, according
to McGaugh (1994), indicating that these galaxies evolve very slowly
and form few stars during a Hubble time.  Apparently, they also
lack detectable gradients.
This, despite the fact that
these objects are similar in mass and size to prominent spirals
defining the Hubble sequence.
McGaugh suggests that a galaxy's environment and surface
mass density are more relevant to galaxy evolution than gross size.

{\it Effects of cluster environment} on the chemical evolution of
galaxies have been investigated by Skillman et al. (1996), who
studied oxygen profiles in several Virgo spirals representing a
range in H~I deficiency (taken as a gauge of cluster environmental
interactions). Their results imply that global metal abundances in
disks tend to be higher in stripped galaxies, presumably because
reduced infall of metal-poor H~I gas means less dilution of disk
material.  Henry et al. (1996 and references therein) investigated
metallicity and heavy element abundance ratios (N/O, S/O) in three
cluster spiral disks with normal H~I and found no clear signatures
of environmental effects.  Thus, cluster environment alone is
apparently not a sufficient condition for altered chemical
evolution.

{\it The mathematical form of abundance profiles in spiral disks}
has been investigated recently by Henry \& Howard (1995), who fit
line strength behavior over the disks of M33, M81, and M101 using
photoionization models. Their best fits for O/H versus
galactocentric distance were produced using exponential profiles,
although power law forms could not be ruled out.  However, linear
profiles poorly reproduced the observations.  Henry and Howard
also concluded that, despite some observational and theoretical
claims to the contrary (see Moll{\'a} et al. 1996),
it is premature to conclude that
gradient flattening is present in the outer parts of some disks.

{\it Planetary nebulae have been used as probes of interstellar
abundances in a small number of external galaxies.} A recent paper by
Jacoby \& Ciardullo (1999) presents abundances for 12 bulge and three
disk planetaries in M31. Their bulge objects have oxygen abundances
whose average is similar to the value observed in the Large Magellanic
Cloud. Interestingly, the implied bulge metallicity is significantly
below the level expected from observations of [Fe/H]. In another study,
Stasi{\'n}ska, Richer, \& Mc~Call (1998) determine abundances of
oxygen, neon, and nitrogen in planetaries in the bulges of the Milky
Way and M31, M32, and the Magellanic Clouds. These authors find higher
oxygen levels in the Milky Way and M31 bulges than in the Clouds, and
also reconfirm the tight correlation between neon and oxygen discussed
below in {\S}2.2.3.

\subsubsection{Summary Thoughts About Spiral Metallicities}

A detailed synthesis based upon the observations is beyond the
scope of our review. However, the following would seem to
provide a reasonable set of explanations.

There appear to be two fundamental physical parameters for a galaxy which
influence its abundance characteristics. These are total mass and the
distribution of material as a function of galactocentric distance. As
supernovae erupt, their metal-rich ejecta are more likely retained in
systems with greater mass. Thus, the more massive galaxies might be expected to exhibit higher global metallicities, which in fact they
do. Furthermore, observations indicate that metallicities tend to be greater in regions where the
total surface density is higher, perhaps because the star formation
process is sensitive to the local density and so more metals are
produced in locations with high densities. Since matter in spirals tends
to form an exponential disk (Binney \& Merrifield 1998) with surface
density falling off with greater galactocentric distance, we might
then expect metallicity locally in the disk to track this pattern.

\subsection{Heavy Element Abundance Ratios In Spiral Disks}

Ratios of heavy elements, i.e. N/O and C/O,
are expected to reveal in particular the
characteristics of the initial mass
function, stellar
yields, and the history of star formation.
Here we consider five ratios which are accessible through
nebular studies,
N/O, C/O, Ne/O, S/O, and Ar/O. Note that because planetary
nebulae are self-contaminating with nitrogen
and (sometimes) carbon, they do not make good probes of the
interstellar levels for these elements, although in the cases of
O, Ne, S, and Ar they seem to work satisfactorily in that
capacity.

\subsubsection{N/O}

We consider the nitrogen abundance studies for the Milky Way
disk indicated in Table~1 along with
H~II region studies by Kobulnicky
\& Skillman (1996), van~Zee et al. (1998), Thurston, Edmunds, \&
Henry (1996), and Izotov \& Thuan (1999) for external spirals.
Figure 6 shows log(N/O) versus 12+log(O/H) for both the Milky Way
and extragalactic objects. Symbols are explained in the caption.

The most striking feature in Fig.~6 is the apparent threshold
running from the lower left to upper right beginning around
12+log(O/H)=8.25 and breached by only a few objects. Behind this
line the frequency of objects drops off toward lower values of
12+log(O/H) and higher values of log(N/O). A second feature is
the behavior of N/O at values of 12+log(O/H)$<$8, where N/O
appears constant, a trend which seems to be reinforced by the upper
limits provided by the damped Ly$\alpha$ objects of Lu et al. (1996; L) at very low
metallicity. This bi-modal behavior of N/O was pointed out
by Kobulnicky \& Skillman (1996).
Although detailed theoretical interpretations are beyond our scope,
we summarize below the basic ideas of nitrogen production and
attempt to tie them to Fig.~6. Readers interested in additional
detail are urged to refer to Vila-Costas \& Edmunds (1993).

Nitrogen is mainly produced in the six steps of the CN branch of the
CNO bi-cycle within H burning stellar zones, where $^{12}$C serves as
the reaction catalyst (see a textbook like Clayton 1983 or Cowley 1995
for nucleosynthesis review).  Three reactions occur to transform
$^{12}$C to $^{14}$N: $^{12}$C(p,$\gamma$)$^{13}$N($\beta ^+
\nu$)$^{13}$C(p,$\gamma$)$^{14}$N, while the next step, 
$^{14}$N(p,$\gamma$)O$^{15}$, depletes nitrogen and has a
relatively low cross-section. The final two reactions in the
cycle transform $^{15}$O to $^{12}$C. Since the fourth reaction
runs much slower than the others, 
the cycle achieves equilibrium only when $^{14}$N accumulates to high
levels, and so one effect of the CN
cycle is to convert $^{12}$C to $^{14}$N. The real issue in
nitrogen evolution is to
discover the source of the carbon which catalyzes the process.

Since stars produce their own carbon during He burning, nitrogen
originating from it is termed {\it primary} nitrogen. Any nitrogen
produced during supernova explosive nucleosynthesis is also termed
{\it primary} since it is created for the first time during the
explosion. On the other hand, stars beyond the first generation in a
galactic system already contain some carbon inherited from the
interstellar medium out of which they formed. Nitrogen produced from
this carbon is termed {\it secondary} nitrogen.

As a system begins to mature chemically from a state of low
metallicity, nitrogen must come from carbon produced by the star
itself, since at this point no significant level of carbon
exists in the ISM which can be incorporated into new
stars and enter into the CN cycle. So, nitrogen production is
primary and its evolution
proceeds at a rate set only by star formation coupled with the
primary production rate of nitrogen. Since the production of
elements such as oxygen is being influenced by similar factors,
the N/O ratio should remain constant as their abundances rise
together.

But as metallicity rises and stars form out of progressively more metal-rich
environments, the amount of carbon present in the star at birth which can
ultimately enter the CN cycle becomes comparable to the amount produced
internally through He burning, and
thus nitrogen production becomes secondary and coupled to the
metallicity of the star.
At this point, N/O versus O/H assumes a positive slope, since
the relation between N and O is now quadratic (Vila-Costas \&
Edmunds 1993).

Based upon the data and models presented in Fig.~6 and allowing
for the scatter, a reasonable explanation for
the observed trend for N/O is that the flatter behavior seen at
12+log(O/H)$<$8.0 corresponds to the dominance of primary
nitrogen production, while the steeper slope in N/O at higher metallcities is
linked to metallicity-sensitive secondary nitrogen
production. We concur
with Shields, Skillman, \& Kennicutt (1991), who found that the
point at which secondary nitrogen production becomes
important is located at roughly
12+log(O/H)=8.3 or 0.6~dex below solar.

A comparison of nitrogen yields from
intermediate mass stars (1-8M$_{\sun}$) by van den Hoek \&
Groenewegen (1997) with those from massive stars by Nomoto et
al. (1997b) suggests that intermediate mass stars are ultimately the main
contributors to nitrogen production, although early-on massive
stars may play a role, due to the longer evolutionary time
scales for less massive stars, and thus their delay in
depositing nitrogen into the interstellar medium.
Vila-Costas and Edmunds (1993) calculated analytical models for
the evolution of N/O assuming a simple, closed-box regime but
accounting separately for primary and secondary nitrogen yields
along with time delays in intermediate-mass star nucleosynthesis.
The two curves in Fig.~6 represent their results using their Eq.~A5
along with two different values for the ratio of time delay
to the system age. We have adopted values for constants
$a$ and $b$ in their formula of 0.025 and 120, respectively, to
force a better fit to the data presented here. The curve
representing the small delay clearly matches the low metallicity
data better, while the curve for greater delay seems to rise
faster at high metallicity and thus fit the data there better. 

Further study of the origin of nitrogen will
require especially more abundances for systems of low metallicity
where 12+log(O/H)$<$7. Studies of damped Lyman-$\alpha$ systems
currently offer great promise in this regard.

\subsubsection{C/O}

Carbon is produced during core and shell helium burning in the
triple alpha process, $3 ^4He \to ^{12}C$.
It is an element whose abundance
has lately become more
measurable in extra-galactic H~II regions,
thanks to the Hubble Space Telescope (HST) and its UV capabilities, since the
strong carbon lines of C~III] and C~IV appear in that spectral
region. Recent studies of extragalactic H~II regions have been
carried out by Garnett et al. (1995; 1997; 1999) and Kobulnicky
\& Skillman (1998), while carbon abundances for M8 and the Orion
Nebula, both within the MWG,
have been measured by Peimbert et al. (1993) and Esteban
et al. (1998), respectively.

Results of these measurements are collected together in Fig.~7,
where log(C/O) is plotted against 12+log(O/H). The point
for Orion is indicated with an `O', M8 with `M', and the sun
with an `S' (Grevesse et al. 1996). The vertical lines connect
points corresponding to carbon abundances determined with two
different reddening laws by Garnett et al. (1998). The filled
circles correspond to stellar data from Gustafsson et al. (1999)
for a sample of F and G stars.

A direct correlation between C/O
and O/H is strongly suggested and has been noted before (c.f.
Garnett et al. 1999),
although the result is weakened somewhat by the two points
for I~Zw~18 around 12+log(O/H)=7.25. Ignoring these two points as well as the ones for the sun and stellar data,
and performing a regression analysis, we find that $log(C/O)=-5.34(\pm 0.68)+0.59(\pm 0.08)[log(O/H)+12]$ (solid line in Fig.~7) with a correlation coefficient of 0.88
when we exclude Garnett et al.'s (1999) data points
corresponding to R$_v$=5 (the connected points with lower C/O),
$log(C/O)=-4.45(\pm 0.60)+0.48(\pm 0.07)[log(O/H)+12]$ (dashed line in Fig.~7) with a correlation coefficient of 0.86 when points for R$_v$=3.1 (the
connected points with higher C/O) are excluded. Assuming that with
additional data the trend becomes more robust, it clearly
implies that carbon production is favored by higher
metallicities.
One promising explanation (Prantzos, Vangioni-Flam, \& Chauveau 1994; Gustafsson et al. 1999) is that mass loss
in massive stars is enhanced by the presence of metals in their
atmospheres which increase the UV cross-section to stellar
radiation. Stellar yield calculations by Maeder (1992) appear to support this
claim. The contributions to carbon by different stellar mass
ranges is discussed by both Prantzos et al. and  Gustafsson et al., who conclude that the
massive stars are primarily responsible for carbon production.
It is also clear, however, that stars of mass less than about
5M$_{\odot}$ produce and expel carbon as well (van den Hoek \&
Groenewegen 1997), and thus the
relative significance of massive and intermediate mass stars is
still not understood completely.

\subsubsection{Ne/O, S/O, \& Ar/O}

Neon is produced through carbon burning
($^{12}$C+$^{12}$C$\to$$^{20}$Ne+$^4$He), 
while both sulfur and argon originate from
explosive oxygen burning in Type~II supernova events 
($^{16}$O+$^{16}$O$\to$$^{28}$Si+$^{4}$He, then $^{28}$Si+$^{4}$He$\to$$^{32}$S;
$^{32}$S+$^{4}$He$\to$$^{36}$Ar). In
addition, substantial amounts of
S and Ar may be manufactured in Type~Ia supernova
events (Nomoto et al. 1997a). Note that here we refer only to the dominant isotopes of the respective elements.

Abundance ratios of Ne/O, S/O, and Ar/O are plotted
logarithmically against 12+log(O/H) in Fig.~8. To the data of Shaver et al.
(1983) and Maciel \& K{\"o}ppen (1994) for the Milky Way we
have added data for Ne/O, S/O, and Ar/O from optical studies of
extragalactic H~II regions
from van~Zee et al. (1998) and Izotov \& Thuan (1999)
along with S/O results from  Garnett (1989) for both the MWG and
extragalactic H~II regions. Representative uncertainties are
$\pm$0.20~dex in each of the three ratios and $\pm$0.20~dex in
12+log(O/H).  The horizontal lines in each panel represent
the predictions from Nomoto et al. (1997a; dashed lines), Woosley \& Weaver (1995; dot-dashed lines), and Samland (1998; solid lines) for massive star yields
integrated over a Salpeter initial mass function between 10-50~M$_{\sun}$ and corrected to give
ratios by number.

All three panels of Fig.~8 show vertical ranges which are
consistent with uncertainties and thus imply in each case a
constant value for each ratio over the 1.5-2 decades of
oxygen abundance. Logarithmic values for unweighted arithmetic
averages (log average antilog) and standard deviations (not
uncertainties) for Ne/O, S/O, and Ar/O are presented in Table~3,
where the first column identifies the sample by the last name of
the first author followed by three numbers indicating the sample
sizes for Ne/O, S/O, and Ar/O, respectively. Also included are
averages for the total of all samples, solar values (Grevesse et
al. 1996), and ratios
found in the Orion Nebula (Esteban et al. 1998) and the Helix
Nebula (Henry, Kwitter, \& Dufour 1999), a nearby planetary
nebula. Generally, for each abundance ratio
there is very good agreement among the
five samples, considering observational
uncertainties. Notice the smaller dispersion
associated with the Izotov \& Thuan data. This may be
explained by their focus
on metal-poor H~II regions possessing very bright emission lines with
resulting signal-to-noise of 20-40 in the
continuum and
abundances frequently having uncertainties of less than $\pm$0.10~dex
compared with the typical $\pm$0.20~dex uncertainties in other
samples (Izotov, private communication). In
addition, we note the significant disparity between Ar/O for Orion
and the other samples along with the sun and the Helix Nebula.
Due to the limited
spectroscopic range, the argon abundance in
Orion was determined using the weak 5192~{\AA} auroral line of
Ar$^{+2}$, where its strength was observed to be on the order of
10$^{-3}$ times H$\beta$. Use
of stronger near IR lines may bring the argon abundance in Orion
into agreement with other objects (Esteban, private
communication). In addition, the S/O ratio found in the Helix Nebula is an order of magnitude below the average value. This is currently difficult to interpret, although a few planetary nebulae do show sulfur abundances which are this low (see Henry et al. 1999).

The evidence provided by Table~3 and Fig.~8 supports the
contention that abundances of Ne, S, Ar, and O evolve in lockstep, a point made by Henry (1989) in his earlier study of Ne and O in planetary nebulae.
This would be expected if these elements are all either produced by
massive stars 
within a narrow mass range or stars of different masses but with an
invariant initial mass function. Under these conditions their
buildup is expected to proceed
in lockstep, and the ratios of Ne/O, S/O, and Ar/O should have
constant values over a range of O/H.

Interesting departures from the
universality of the ratios displayed in Fig.~8
appear in halo planetary nebula studies. The
detailed one by Howard, Henry, \& McCartney (1997), for example,
confirms earlier indications that the object BB-1 has
log(Ne/O) of -0.11, while log(Ne/O) for H4-1 has a value of
-1.82. These deviants might be explained by local abundance
fluctuations caused by recent supernova events whose ejecta,
differing in composition because of mass cut differences in the
explosive event, had not yet mixed in with the surrounding ISM
before the PN progenitor formed out of it.

Finally, notice that predicted ratios from the yields for stars in the
10-50~M$_{\odot}$ mass range, represented by the horizontal lines, generally fall below the observed average, with the offset for Nomoto et al. consistently being the largest. This suggests that the theoretical calculations
overproduce oxygen and further imply that the adopted
rate of the $^{12}$C($\alpha,\gamma$)$^{16}$O reaction in the models is too
high, resulting in a higher conversion rate of $^{12}$C to
$^{16}$O with a boost in the oxygen production relative to
elements such as Ne, S, and Ar. This conclusion agrees at least
qualitatively with comparisons by Nomoto et al. (1997b) of yields
from two 25~M$_{\odot}$ stellar models using significantly different
values of the $^{12}$C($\alpha,\gamma$)$^{16}$O rate. Another
explanation in the case of S/O and Ar/O may be that the predicted yields
do not include contributions
from Type~Ia
supernovae, which produce significant amounts of $^{32}$S and $^{36}$Ar,
according to Nomoto et al.'s (1997a) W7 model. Adding this source
to yields of massive stars would raise the theoretical line.
This subject should be explored in more detail, particularly
since the Ne/O ratio ought to provide a good constraint on the
value of the $^{12}$C($\alpha,\gamma$)$^{16}$O rate.

\section{Abundance Patterns In Galaxies From Stars}

Stars are substantially fainter than H II regions, planetary nebulae,
or supernova remnants in general,
and therefore can only be observed one by one in
the Milky Way except for the brightest giants and supergiants, some of
which can be observed out to about 10 Mpc. Additionally, abundance
measurements in stars come from relatively precise measurements of the
depth of absorption features (rather than emission features),
therefore requiring more photons for results of similar
accuracy. Furthermore, it is usually the more inconspicuous lines from
which the most reliable abundances are derived! However, stars have
one distinct astrophysical advantage: they are long-lived. Study of
stars of different ages can reveal chemical history explicitly, rather
than implicitly through chemical evolution models. In this
section we briefly survey the extant stellar results for the MWG
and external galaxies. A brief description of how abundances are
inferred from stellar spectra is provided in Appendix~B.

\subsection{Milky Way Galaxy}

To summarize four decades of work on Milky Way stellar abundances in a
balanced fashion is clearly beyond the scope of a single paper, so we
attempt to provide an executive summary. Workers now subdivide the
Milky Way into the spheroidal halo ($r>2$ kpc) and bulge ($r <2$ kpc)
and two disk-like components, the thick disk (scale height $\approx 1$
kpc) and the thin disk (scale height $\approx 350$ pc). We now
treat these in order.

The halo has a metal abundance of [Fe/H] $\approx -1.6 \pm
1$\footnote{Here we employ the standard bracket notation often used in
expressing abundances, $[X] \equiv log(X)-log(X)_{\sun}$, where
X represents an elemental abundance or an abundance ratio. We will use
the symbol $Z$ to represent all heavy elements at once. } with no
noticeable abundance gradient. This information comes from studies of
individual subdwarfs (Carney et al. 1990, 1996). Globular clusters
fall into two spatially and kinematically distinct groups; the inner,
metal-rich disk clusters and the outer, metal-poor halo
clusters. Internal to either group there is no clear abundance
gradient (Zinn 1996; Richer et al. 1996). Increasing
evidence suggests that the globular cluster system has a significant
age spread of 3-4 Gyr (e.g. Hesser et al. 1997), which may depend on
radius; older toward the center. There is also a pattern of lighter
species becoming enhanced relative to Fe-peak elements in stars more
metal-poor than about [Fe/H] = $-1$, with O, Mg, Al, Si, Ca, and Ti
overabundant by several tenths of a dex relative to a scaled-solar
mixture in the metal-poor group (Wheeler et al. 1989; Edvardsson et
al. 1993). The standard interpretation is that the metal-poor group
was enriched mainly by Type II supernova nucleosynthesis products and
the metal-rich stars contain a mixture of Type II and Type I products,
where the Type I supernovae are thought to produce mainly Fe-peak
elements.
 
The Galactic bulge has much foreground dust as well as confusion with
foreground disk stars and thus is a difficult place for observational
work.  Photometric studies seem to indicate a negative
abundance gradient but the size of the gradient is not yet
well-quantified (Terndrup 1988; Frogel et al. 1990; Harding
1996). Due to the reddening, spectroscopy seems like a safer way to
proceed. Still, results are ambiguous. Ibata \& Gilmore (1995) find a
near-solar metallicity but no gradient outwards from 0.6~kpc, but
Terndrup et al. (1990) and Rich (1998) find a gradient of about $-0.4$
dex/kpc ( or $\Delta$~log($Z$)/$\Delta$~log~$R \approx -0.6$ )
considering regions somewhat further toward the Galactic center.

The thick disk, massing about 10\% of the thin disk, is separable
from the thin disk and halo primarily through kinematics or age, since
its metallicity overlaps at the high end with the thin disk (Wyse \&
Gilmore 1995) and at the low end with the halo (Nissen \& Schuster
1997). No radial or vertical gradient in [Fe/H] has been discovered in
several large data sets (Gilmore et al. 1995; Bell 1996; Robin et
al. 1996). The high-quality data of Edvardsson et al. (1993) confirms
the lack of a {\em strong} gradient for heavy elements, but finds a
probable relation between ``alpha'' elements Si and Ca relative to Fe
as a function of radius over a 4- to 12-kpc span of about
[$\alpha$/Fe]/$R_m =$ +0.03 dex/kpc, where $R_m$ is an estimate of the
radius at which the stars were born, rather than where they are
presently located. Similarly, [$\alpha$/Fe] has been found to increase
with age.  Small amplitude results of this nature should become more
common as stellar abundances become more accurate.

The thin disk can be traced by open clusters, most of which are
younger than about half the age of the disk. Friel \& Janes (1993) and
Thogersen et al. (1993) worked with moderate resolution spectra of
open cluster K giants to obtain a mean [Fe/H] gradient of $-0.097
\pm 0.017 $ dex/kpc between 7 and 15 kpc. Photometry of open cluster stars
has yielded similar results (Panagia \& Tosi 1981; Cameron 1985). An
alternative to a steady gradient has been proposed by Twarog et
al. (1997), who from a sample of 76 open clusters find a sharp
falloff of roughly 0.35 dex at around 10~kpc from the Galactic
center with flat gradients interior and exterior to that radius.

Of similar luminosity to K giants are B main sequence stars. B stars
are youthful in age, so their abundances should match those of H II
regions. Accurate spectral analysis of oxygen lines in (at least) the
hotter B stars is dependent on dropping the assumption of ``local
thermodynamic equilibrium'' between the radiation field and the
gas. Smartt \& Rolleston (1997) and Gummersbach et al. (1998) have undertaken such an analysis,
deriving an [O/H] gradient of $-0.07 \pm 0.01$ dex/kpc, a result
different from most of its predecessors, but in agreement with the
nebular results.

Figure 9 shows a summary of the broad-brush abundance pattern in the
Milky Way. The metal-poor halo weighs about $10^9 M_\odot$
(e.g. Freeman 1996) compared to about $60\times 10^9 M_\odot$ for the
total mass (in stars) in the Galaxy, so the number of symbols on the
plot does not reflect where the mass is; most of the mass resides
in the disk. Two trends are evident from the figure: abundance
increases with time, and the abundance is higher toward the Galactic
center. Coupled with stellar kinematics and age information, these
abundances give a picture of Galaxy formation in which the halo formed
early and without much chemical enrichment. The disk may have started
early as well, but it is still gas rich and is still forming stars
today at near-solar abundance.

\subsection{External Galaxies}

Spectroscopy of individual stars in local group galaxies M31 and M33
has become possible in recent years for supergiants, typically
A-type. Like B main sequence stars, A supergiants suffer from serious
non-LTE effects in the outer photosphere, but lines can be chosen that
form deep in the photosphere and a partial non-LTE analysis can be
attempted for other interesting lines. The resultant accuracy can be
$\pm 0.2$ dex (Venn 1995; 1998). For M31, the [O/H] gradient obtained
from A supergiants is consistent within the errors with that obtained
from nebular studies (McCarthy et al. 1998). For M33, based on four B
supergiants, Monteverde et al. (1997) obtain an [O/H] gradient of
$-0.16 \pm 0.06$ dex/kpc, which is also similar to nebular results.

Most stellar abundance work in external galaxies relies on the colors
of red giant stars from older populations. After spectroscopic
abundance work in globular clusters showed a wide range of
metallicities among clusters, it was obvious from published
color-magnitude diagrams (CMDs) that the red giant branches are redder
at progressively higher metallicities. This finding can be used as an
abundance indicator, especially in the HST era where the tip of the
red giant branch can be seen at distances of $\sim 10$ Mpc.  A younger
age population has a somewhat bluer giant branch, but this effect is
fairly minimal, and in some cases negligible when the age is already
known. Crowding of stars excludes near-nuclear regions from CMD
analysis.

The halo of M31 has been examined by Durrell et al. (1994) and Rich et
al. (1996) from HST optical colors, with the conclusion that, like the
Milky Way, no abundance gradient is apparent. Unlike the Milky Way,
the average abundance of the stars is [Fe/H] $\approx -0.6$ (Durrell
et al.) or even higher (Rich et al.). Grillmair et al. (1996) derive
an abundance distribution (number of stars per interval [Fe/H]) for
the outer disk of M31 that is identical within the errors with the
abundance distribution of the solar neighborhood. Elliptical galaxies
NGC 5128 (Soria et al. 1996) and M32 (Grillmair et al. 1996) have also
been studied in this fashion, but only at a single radius so far, so
we await further data before we can draw conclusions about abundance
profiles.

\section{Abundance Patterns In Spheroidal Systems From
Photometric Observations}

\subsection{Metallicity Gradients}

We now provide an overview of the abundance profile picture for
spheroidal systems, especially elliptical galaxies. The
principal techniques for measuring abundances in these systems
involve the use of photometric indices of integrated starlight,
since individual stars cannot be resolved. A brief description
of these techniques is given in Appendix~C. Elliptical galaxies look
superficially like a fairly homogeneous class of galaxies, with muted
star formation and no obvious cold gas, and kinematically supported by
almost randomly oriented orbits. Star formation can be seen in most
ellipticals at some level, as can dust lanes and emission-line gas,
but usually at a level below that of spirals. Ellipticals also exhibit
regularity of colors and absorption feature strengths, with larger
galaxies being redder and having stronger metallic absorption features
than smaller ones. This has long been interpreted as a sign that
the metallicity is higher in larger galaxies (e.g. Faber 1972).

To derive abundance profiles in stellar systems, colors and absorption
feature strengths as a function of galactocentric radius are
interpreted through population models. Some color gradient studies
include Kormendy and Djorgovski (1989), Franx \& Illingworth (1990),
and Peletier et al. (1990). Most studies of optical absorption
features have utilized one particular system of feature indices
developed at Lick Observatory (described in Worthey et al. 1994 and
references therein). The last few years have seen a rapid expansion of
galaxy data available in this system. To measure an absorption feature
in the Lick system, one creates a pseudocontinuum by bracketing the
spectral feature of interest with flanking passbands. Flux in the
flanking bands is measured and a straight line is drawn between the
midpoints of the flanking bands to represent the
(pseudo)continuum. The flux difference between the pseudocontinuum and
the absorption feature is integrated and the result is expressed in
\AA\ of equivalent width (or magnitudes, depending on the specific
index; see Worthey et al. 1994 and Worthey \& Ottaviani 1997 for the
details.) Figure 10 illustrates the idea for a portion of the
spectrum. There are 25 indices defined, 5 definitions measuring Balmer
lines and 20 measuring various metallic absorption blends. The index
system operates at a low resolution ($\sim 8$ \AA\ FWHM) necessitated
by Doppler smearing from the substantial (up to $\sigma = 350$ km/s)
velocity dispersions of large elliptical galaxies, and most of the
indices require corrections when velocity dispersions get large.

While most of these 25 indices follow the 
$\Delta$~log(Age)/$\Delta$~log~$Z \approx -{{3}\over{2}}$ constant-index slope
described in Appendix~C, a few
(the Balmer indices) are relatively age sensitive, with
$\Delta$~log(Age)/$\Delta$~log~$Z \approx -{{1\ {\rm to}\
2}\over{2}}$, while others, notably a feature called Fe4668 whose main
contributor is really molecular carbon, are relatively metal sensitive,
with $\Delta$~log(Age)/$\Delta$~log~$Z \approx -5$ (Worthey 1994).
Arrayed against each other, it seems possible to separate the effects
of age and metallicity, in the mean. 

To derive an abundance gradient in an elliptical galaxy, one compares
observed colors or line strengths with model predictions, often assuming a
constant age throughout the galaxy. For instance, Franx \& Illingworth (1990) find a mean
color gradient in 17 elliptical galaxies of $\Delta(U-R)/\Delta {\rm
log}\ r = -0.23 \pm 0.03$ mag per decade in radius. Entering the
Worthey (1994) models at age 12 Gyr, one finds that a change of 0.15
dex in $Z$ gives the required $\Delta(U-R)$, so the gradient assuming
constant age is $\Delta {\rm log}\ Z/\Delta {\rm log}\ R=-0.15$ dex
per decade. The same number is reached by considering the $B-R$
gradient. Due to the very steep surface brightness dropoff of
elliptical galaxies, projection effects are small and usually
neglected. The steep dropoff also means that long-slit spectroscopy
usually only reaches to 0.5 to 1.0 $R_e$ (the half-light radius)
although color gradient studies and ultradeep spectroscopy can reach
to several $R_e$.

Color studies and line strength studies generally give a consistent
picture of a gradient of about $\Delta$log$Z$/$\Delta$log$R \approx
-0.2$ dex per decade. There is probably a small correction to this
number, however, due to age effects. Simultaneous mean-age, mean-$Z$
estimates using the Balmer-versus-metal feature technique described
above were derived for the Gonz\'alez (1993) and Mehlert et al. (1998)
samples of galaxies, about 60 early type galaxies in a wide variety of
environments, and the distribution of gradients is shown in
Fig. 11a. There are no trends of gradient strength with luminosity or
velocity dispersion (unlike average $Z$, which is larger in larger
galaxies). The average age gradient is younger toward the center by
0.1 dex/decade (a few Gyrs), and more metal-rich by 0.25
dex/decade. The scatter in the average seems mostly due to
observational error, error in correcting for H$\beta$ emission
fill-in, and variation in abundance ratio mixture, and the
residuals scatter along the $-{{3}\over{2}}$ age-metallicity
slope (Fig. 11b) in the way expected for random errors in input index
values. 

This $-0.3$ gradient in dex/decade units corresponds to about $-0.02$
dex/kpc assuming an 8~kpc radius, which is a factor of three more
shallow than the gradient found for the Milky Way disk and other
non-barred spirals (see {\S}2.1).  But such a value is well within the
range of theoretical models for galaxy formation. For example,
Larson's (1974) dissipative models predict
$\Delta$log$Z$/$\Delta$log$R = -1$, while various Carlberg (1984)
models range from $-0.5$ to 0.0. Pure stellar merging gives zero
gradient, and in fact tends to erase pre-existing gradients by roughly
20\% per event, or even more via changes in radial structure of the
galaxies (White 1980).

The gradient numbers seem fairly robust and consistent from data set
to data set and from model to model. What about absolute abundances?
These are trickier. The nuclei of large elliptical galaxies have mean
[Z/H] in the range 0.0 to 0.4 dex as inferred from Balmer-versus-metal
feature diagrams. In principle, the mean abundance can be known 
much more precisely, but there is an additional stumbling block
beyond just the inaccurate models and the complication of age-metal
degeneracy. The elemental mixture in elliptical galaxies is not
scaled-solar. Abundances derived from lighter-element lines like Mg b
or Na D are much higher than those derived from heavier Fe or Ca
lines, and this is the {\it main cause for uncertainty in the absolute
abundance} (Worthey 1998).

\subsection{Enhanced Light-to-Heavy Element Ratios}

The light element\footnote{In this subsection we make a distinction
between ``light'' and ``heavy'' elements, divided at the fourth row of
the periodic table, so that Ca and Fe are heavy, but Na, Mg, and N are
light. The oft-standard terminology is to speak of ``alpha'' elements,
but ``alpha'' usually includes Ca and excludes N, which makes little
sense for the abundance pattern seen in massive elliptical galaxies.}
enhancement can be seen in the case of [Mg/Fe] by plotting a magnesium
feature index (Mg$_2$) versus an average iron feature ($<{\rm Fe}> =$
the arithmetic average of indices Fe5270 and Fe5335).  The age
sensitivities of these indices are about the same ($\approx
{{3}\over{2}}$!), so models of different ages and metallicities should
lie on top of one another. They do, as seen in Figure 12, but at high
Mg strength the galaxies follow another distinct trajectory entirely with some
galaxies trending toward strong Mg$_2$ strength at nearly constant
$<$Fe$>$, and hence, we infer, relatively enhanced Mg abundance.

The models are scaled-solar since they are built from local stars, so
they cannot track altered abundances. Composite populations add
approximately like vectors, so any combination of ages and
metallicities still lands on the same model locus. Different models
built by different authors have a spread of something like $\pm 0.5$
dex at constant index strength, but all models follow almost exactly
the same slope in the Fig. 12 diagrams, so that the inferred
[Mg/Fe] for the high-Mg$_2$ group of elliptical galaxies is in the
range [Mg/Fe] = 0.3 to 0.5 dex. 

A crucial thing to notice is that velocity dispersion tracks Mg$_2$
very tightly, so the high Mg$_2$ galaxies are also the largest
galaxies (or, more precisely, the ``dynamically hottest''). The
Mg$_2$-$\sigma$ relation is shown in Figure 13; it is one of the
tighter scaling relations known, much tighter than, say, the
$<$Fe$>$-$\sigma$ diagram, which is almost a scatter plot.
With the existence of the Mg-$\sigma$ relation,
[Mg/Fe] increases with galaxy size, where a velocity dispersion of
$\sigma \approx 200$ km/s seems to mark the beginning of noticeable Mg
enhancement.

Figure 12 shows separate diagrams for spiral bulges, S0
galaxies, and elliptical galaxies. No marked difference between Hubble
types is seen except that already ascribed to velocity
dispersion. That is, spiral bulges hover near the solar ratio area,
but only two bulges have $\sigma > 200$ km/s (and those are on the
high-Mg side of the distribution). Elliptical galaxies possess both the
most extreme velocity dispersions and the most extreme Mg enhancement.

The gradient vectors shown in Figure 12 tend to parallel the
age-metallicity direction 
traced by the various models rather than the more horizontal slope
defined by the nuclei.
This would suggest that the Mg enhancement is global
throughout the galaxy rather than concentrated only at the nucleus. Is
this a hint that the enrichment mechanism (presumably supernova)
spreads enriched gas 10 or 20 kpc from its origin, or does it merely
imply effective mixing?  

The [Mg/Fe] data suggest a variation in enrichment from Type Ia
(mostly Fe) and Type II (all elements) supernovae passing from small galaxies or
bulges to large ones, in the sense that the large galaxies have more
Mg and hence comparatively more Type II enrichment. Figure 14 shows
some corroborating evidence from Trager et al. (1998) nuclear data in
that Ca appears to track Fe, while Na and N are enhanced in a way
similar to that of Mg; only in the larger galaxies or bulges. 

The mechanism for modulating Type~I/Type~II enrichment is not
known. It could be due to a time delay in Type I metal production, or
could be some other connection to velocity dispersion like a variable
IMF or a variable fraction of binary stars. (Worthey, Faber, \&
Gonz\'alez 1992; Weiss, Peletier, \& Matteucci 1995).

\subsection{The G Dwarf Problem: a peaked abundance distribution}

The abundance distribution (number of stars versus [Fe/H]) in the
Milky Way galaxy is more strongly peaked than the simplest closed-box
model (see {\S}1) with constant yield predicts. This is known as the G dwarf
problem (van den Bergh 1962; Pagel 1997, Cowley 1995). It seems almost
certain now that other galaxies share this ``problem'' of having a
relatively peaked abundance distribution. Part of the evidence comes
from integrated light (Bressan et al. 1994; Worthey, Dorman, \& Jones
1996) via three lines of evidence.

First, around 2600 \AA\ there is a paucity of ultraviolet flux which
would otherwise be greater if large numbers of metal-poor main
sequence stars are present. Second, in small compact ellipticals for
which data exist, a high-resolution Ca II index (Rose 1994 system, not Lick
system) detects few A-type horizontal branch stars, where
more of these objects would be expected if a large metal-poor
population exists. Third, if a metal-sensitive index like Fe4668 is
modeled using the broad simple model predictions, the strong line
strengths in large galaxies are very difficult to attain, and require
improbably high yield values. Clinching the integrated light results,
recent color-magnitude diagram studies of individual red giants in the
compact elliptical M32 (Grillmair et al. 1996), large elliptical NGC
5128 (Soria et al. 1996), and the disk of M31 (Grillmair et al. 1996)
all indicate a very peaked abundance distribution similar to or more
peaked than that of the Milky Way.

These empirical findings represent important constraints
on some galaxy formation and chemical evolution
models, and their implications are only starting to be explored (Larson
1998).

\subsection{Assorted Issues}

{\it The Mg$_2$-$\sigma$ relation} (Fig. 13) is tighter than other
population-to-structural correlations such as Mg$_2$-$M_B$, or $<{\rm
Fe}>$-$\sigma$. (Bender, Burstein, Faber 1993) There is some powerful
connection between velocity dispersion and Mg abundance 
as traced by the Mg b feature,
the exact nature of
which eludes us at the moment. One possibility, suggested by Faber et
al. (1992), is that cloud-cloud collision velocity modulates the IMF
to favor more massive stars in high-$\sigma$ environments. This
suggestion is in harmony with the Mg/Fe abundance trend. Another
possibility is that larger local escape velocities resist supernova
winds more effectively, holding onto heavy-element contaminants
better. Star formation is finally truncated when supernova winds are
able to blow the remaining gas out of the galaxy. This is the
now-standard picture of chemical evolution in elliptical galaxies
(e.g. Arimoto \& Yoshii 1987, Matteucci \& Tornamb\'e 1987). This
picture can also be made harmonious with the Mg/Fe trend if there is
an additional mechanism for varying Mg/Fe as a function of galaxy
size, but the fact that Fe abundance is virtually independent of
galaxy size causes some trouble.

{\it Discontinuities in kinematic profiles are coincident with
discontinuities in line strength profiles.} Bender and Surma (1993)
discovered in the course of studying peculiar kinematics in elliptical
galaxies that many have counter-rotating cores or other kinematic
discontinuities. In every case that they studied, a discontinuity
appeared in the Mg$_2$ line strength profile at the same location as
the kinematical discontinuity. At the very least, this implies that
whatever formation mechanism produced the distinct core also
influenced the local chemistry. Muted echoes of formation exist still
in both the stellar kinematics and the chemical signature in the
stars.

Study of the {\it globular cluster systems around elliptical galaxies}
yields insight into the formation of halos in general and elliptical
galaxies in particular. Some elliptical galaxies host a very large
number of globular clusters per unit luminosity (e.g. M87) while
others have about as many as are seen in spiral galaxies. Questions
remain about how the globular clusters are created and destroyed to
explain the wide variation in number and whether merging events are
important or not (van den Bergh 1995; Zepf \&
Ashman 1993). Most abundance studies concentrate on the integrated
colors of the clusters (Ostrov et al. 1998; Lee et al. 1998; Ajhar et
al. 1994) because of their faintness, but some spectroscopic studies
are beginning to appear (e.g. Cohen et al. 1998). These efforts show a
variety of interesting results. For example,
some globular cluster systems are metal-poor, some
metal-rich, and some bimodal or multimodal. And when both metal-poor
(blue) and metal-rich (red) populations coexist, the red population
tends to be more centrally concentrated than the blue. 

\section{Summary and Suggestions}

We have explored in some detail the abundance patterns in spiral disks and elliptical galaxies as revealed through analyses of gaseous nebulae, stars and integrated photometry of galaxies.

The principal points regarding abundance patterns in spiral disks
are:

\begin{itemize}

\item The metallicity as gauged by O/H in nebulae across the Milky Way disk decreases with galactocentric distance, a finding supported by recent abundance results for disk stars. This negative gradient pattern is seen in most other spiral disks. A similar result is seen when luminous stars are used as abundance probes. Scatter at any particular galactocentric distance is consistent with observational uncertainty.

\item Global metallcity, taken as the abundance of oxygen at a standard galactocentric distance, is positively correlated with galaxy mass.

\item Metallicity at any location in a spiral disk appears to be positively correlated with the local total surface density.

\item Abundance gradients are steeper in normal spirals than in barred ones.

\item A plot of N/O versus O/H in spiral disks indicates that production of nitrogen is dominated by primary processes at low metallicities and secondary processes at high metallicities.

\item C/O is positively correlated with O/H in spiral disks, suggesting that
carbon production is sensitive to metallicity, possibly through metallicity-enhanced mass loss in massive stars.

\item Abundance ratios of Ne/O, S/O, and Ar/O appear to be universally constant across the range in metallicities observed, reflecting the idea that either the initial mass function is universally constant; or the stellar mass range responsible for producing these elements is relatively narrow, and thus these ratios are insensitive to IMF variations.

\item Stellar age and galactocentric distance in the Milky Way show rough correlations with metallicity in the sense that metallicity
decreases with increasing age and galactocentric distance. However, all
Galactic components (halo, bulge, thin disk, thick disk) have large scatter
in abundance, and even the metal-poor halo is now thought to display
age scatter of several Gyr.

\end{itemize}

For elliptical galaxies, the main results are:

\begin{itemize}

\item Abundance gradients
are, on average, about a factor of two to three
more shallow than in non-barred spirals. This is well within the range
expected from various formation pictures, including hierarchical
mergers of smaller galaxies.

\item Nuclear or global metallic feature strengths (or colors) become
stronger (or redder) in larger galaxies. The 1970's conclusion that
larger elliptical galaxies must be more metal-rich is reconfirmed, but
every elemental species does not increase in lockstep.

\item Light elements N, Na, and Mg are enhanced relative to heavy
elements Ca and Fe in the largest elliptical galaxies, implying a
modulation of enrichment, plausibly due to variance of the Type II to
Type Ia supernova ejecta, compared to smaller ellipticals, bulges, and
disks.

\item The mean abundance near the nuclei of large elliptical and S0
galaxies is uncertain, but is in the range [Z/H] = 0.0 to 0.4. Most of
the difference in abundance between small and large galaxies is driven
by the increasing abundance of elements {\it lighter} than those near the
Fe-peak, with [Fe/H] staying roughly constant for elliptical galaxies
of all sizes.

\item The
abundance distribution in elliptical galaxies and, so far, every other
well-studied large galaxy type, is strongly peaked like that of the solar
cylinder, not broad like the simplest closed-box model predicts.

\end{itemize}

There are two over-arching patterns which emerge from the combined
results for spirals and ellipticals. First, {\it there is a positive
correlation between galactic metallicity and mass}.  This may be
related to the greater retension of heavy elements ejected by
supernovae by the stronger gravitational potentials of massive
galaxies, or perhaps to the effects of galaxy mass on the star
formation process. It is currently difficult to ascertain whether this
relation is completely continuous across galaxy types; in other words,
if one plotted global metallicity versus mass for a sample of galaxies
containing both spirals and ellipticals would there be an unbroken
straight line, or would the correlation for one type be offset from the
other. The difficulty here is in directly comparing abundances between the two
galaxy types. As we have seen, metallicity in spirals is generally
gauged by observing oxygen in nebulae located in their disks. Yet in
ellipticals, metallicity must be measured from integrated light using
numerous photometric indices which are affected not only by
metallicity but by age. Thus, no seamless technique exists for
determining abundances consistently for spirals and ellipticals, and
thus intercomparisons are problematic. This is made all the more
complicated by the fact that we currently don't know how to represent
the global abundance in a galaxy. Do we take the abundance at the
nucleus, or at one effective radius, or at 0.4 optical radii?

The second pattern which has emerged is that {\it abundance gradients
appear to become flatter as one progresses from normal spirals to
barred spirals to ellipticals}. The difference between normal and
barred spirals is currently explained by enhanced radial gas flows in
the disks of barred spirals. To
extend this model to ellipticals it may be neccessary to invoke other
radial mixing mechanisms, either during primordial formation or during
later merging events.
If the pattern is discontinuous between spirals
and ellipticals this might suggest that different processes operate
to affect the gradients in the two galaxy types. Again, our lack of
ability to intercompare spiral and elliptical abundances prevents
further exploration of this pattern at present.

Understanding the broad picture of galactic chemical evolution will
require us to firm up the links between elliptical and spiral galaxy
abundances. While a common elemental yardstick may not exist because of the
different elements which we observe directly in each galaxy type, it
may be possible to tie the two types together abundance-wise by
observing elements in each which share the same production site
nucleosynthetically speaking. An example might be oxygen and
magnesium. In external spirals oxygen is taken as the metallicity gauge
primarily because of its observability. Magnesium, which, like oxygen,
is primarily produced in massive stars (Nomoto et al. 1997a,b) may be
measurable directly through a calibrated Mg$_2$ index. Then oxygen and
magnesium might be linked by assuming a ``cosmic'' Mg/O ratio
calibrated locally. Also, although the work is not started, it may be
possible to construct an oxygen-sensitive photometric index for
integrated light, perhaps revolving around the 2.3$\mu$m CO feature in
conjunction with the C$_2$-sensitive 4668 feature. As synthetic
spectra and stellar abundances grow more precise, these speculative
suggestions might take place, leading to a much more clear
understanding of chemical enrichment and galaxy formation.

\acknowledgments

This collaboration was inspired by the October, 1997, workshop
``Abundance Profiles: Diagnostic Tools For Galaxy History'' held at
Universit{\'e} Laval, Qu{\'e}bec. We are grateful to the
organizers for giving all of us the opportunity to come together and share
our ideas about what one participant, in an attempt to relabel the abundance
field with a trendier and more attention-grabbing name, referred to as
``bio-resources''.  We also thank Bill Blair, C{\'e}sar Esteban, Mike
Fich, George Jacoby, Yuri Izotov, Joachim K{\"o}ppen, Walter Maciel,
D\"orte Mehlert, Anne Sansom, Jan Simpson, and Friedl Thielemann for promptly responding
to inquiries with useful answers and information given generously. And finally, we are especially grateful to our referees, Dave Burstein, Karen Kwitter, and Bernard
Pagel, for promptly and carefully reading the manuscript and making
numerous constructive comments which have improved the paper
tremendously.

\appendix

\section*{Appendix A}

\section*{Abundances From Emission Line Objects}

We present here only a brief overview of abundance determining
methods relevant to emission-line objects. A list of prominent nebular emission lines is presented in Table~A1, where we provide the ion, wavelength, and dominent excitation mechanism for each line.  Readers interested in
greater detail are urged to consult information in Spitzer (1978),
Aller (1984), Osterbrock (1988; 1989), and
Williams \& Livio (1995). The atomic
data used for the abundance calculations are reviewed by Butler (1993).

The basic method for obtaining the abundance of an element in emission-line
objects comprises two steps: (1)~determine abundances of the ions of
that element whose emission lines are directly observable; and
(2)~adjust the
total of the ionic abundances by a factor which accounts for
ions of the same element which are unobservable.

Consider step 1.
An ionic abundance relative to H$^+$ is related to
the observed strength of an emission line integrated over
wavelength,
corrected for interstellar
reddening, and expressed relative to the H$\beta$ strength,
I$_{\lambda}$/I$_{H\beta}$, through the respective
reaction rate coefficients, $\epsilon_{\lambda}$ and
$\epsilon_{H\beta}$ in erg cm$^3$ s$^{-1}$ sr$^{-1}$, such that:
\begin{equation}
\eqnum{A1}
{{I_{\lambda}}\over{I_{H\beta}}}={{\int
\epsilon_{\lambda}(T_e,N_e) N_i
N_e ds}\over{\int \epsilon_{H\beta}(T_e,N_e) N_{H^+} N_e ds}}
\end{equation}
where $N_i$, $N_{H^+}$, and $N_e$ are local number densities of the
ion giving rise to the line $\lambda$, H$^+$, and electron
density. The integrals arise because local products are
integrated along the line of sight. Note that the rates are
functions of the local electron temperature (T$_e$) and density
(N$_e$), although a good simplifying assumption is that these
are constant within the nebular regions actually dominated by the ions in
question.
$T_e$ is usually
determined using the line strength ratio of two lines
such as [O~III] $\lambda$4363 and $\lambda$5007 whose upper
energy levels are relatively far apart.
N$_e$ is derived from the ratio of two lines such as
[S~II] $\lambda$6716 and $\lambda$6731 whose upper energy levels
are closely spaced but the transitions differ significantly in
their sensitivities to collisional deexcitation.

Adding the observed ionic abundances for an element together
gives us a subtotal which differs from the desired total by the
abundances of the ions whose emission lines are not observed.
Thus, in step~2 above we determine an ionization correction
factor, icf(X), for element X by which we multiply the subtotal
to produce the total elemental abundance relative to H$^+$.
Mathematically, the number
density of an element $N_X/N_H$ can be expressed as:  
\begin{equation}
\eqnum{A2} 
{{N_X}\over{N_{H^+}}} = \left\{{\sum^{obs}}{{N_i}\over{N_{H^+}}}\right\} \cdot icf(X/H).  
\end{equation}
Ionization correction factors may be inferred from model
simulations of
nebulae or estimated by assuming that ions with
similar ionization potentials are present in the gas in similar
ratios to their total abundances. Thus, because the ionization potentials of O$^{+2}$ and
He$^+$ are 54.9ev and 54.4ev, respectively, the total abundance of
unobservable (in the optical) higher ionization stages such as O$^{+3}$
and beyond with respect to total O is similar to relative amounts of
He$^{+2}$, an optically observable ion, with respect to total He.
A good compilation and discussion of a
broad range of icf's may be found in the appendix of Kingsburgh
\& Barlow (1994). Additionally, model grids such as those by 
Stasi{\'n}ska \& Schaerer (1997) may be used 
to derive icf's for a nebula assuming the central star temperature is known.

The above method breaks down most frequently for metal-rich nebulae
with low equilibrium temperatures, in which case auroral lines
such as [O~III] $\lambda$4363 are too weak to measure, and thus
the temperature cannot be determined. One way
around the problem is to calculate a detailed photoionization model of the
nebula using
input abundances and other physical parameters which produce an
output set of line strengths closely matching the observed ones. Actual abundances are then inferred from the model input.
A simpler solution is the
``strong-line method'', which uses a composite of
strong, observable emission line strengths 
whose value tracks an abundance ratio. The
most significant example is $R_{23} \equiv ([O~III] +
[O~II])/H\beta$,
first introduced by Pagel et al. (1979), which comprises the sum of
[O~III] and [O~II] nebular line strengths relative to the strength of
H$\beta$ and is related in a complicated but understandable way
to the total oxygen abundance O/H.
This method has been refined and discussed more recently by
Edmunds \& Pagel (1984), Edmunds (1989) and McGaugh (1991), and
of course it is not problem-free.
Because the metal-rich portion of
this relation must currently be calibrated with models, uncertainties
arise from parameter choices such as relative 
depletion (Henry 1993; Shields \& Kennicutt 1995)
and gas density (Oey \& Kennicutt 1993).  Finally, analogous methods
for obtaining N/O and S/O are presented by Thurston,
Edmunds, \& Henry (1997) and D{\'i}az (1999), respectively.

Finally, the accuracy of abundances in emission-line systems is threatened
by the
proposed existence of small scale temperature fluctuations along
the line-of-sight, first described by Peimbert (1967).  In this
picture, an electron temperature measured with forbidden lines is
actually overestimated when fluctuations are present but ignored.
This in turn causes an underestimation of an abundance ratio such
as O$^{+2}$/H$^+$ when it's based upon a forbidden/permitted line ratio
such as [O~III] $\lambda$5007/H$\beta$. When ratios of permitted
lines are used the effect is minimal and so abundances inferred
from permitted/permitted line ratios are unaffected and
systematically higher than abundances from forbidden/permitted
ratios. Temperature fluctuations have been used to explain, among
many other things, the significant discrepancy in planetary
nebula carbon abundances (Peimbert, Torres-Peimbert, \& Luridiana 1995), where those determined using C~II
$\lambda$4267/H$\beta$, say, are often several times greater than
abundances inferred from C~III] $\lambda$1909/H$\beta$. Esteban et al. (1998) found the effect to be small in the Orion Nebula, while Liu (1998) found a large effect in the planetary nebula NGC~4361, although it was insufficient for explaining the discrepancy between carbon abundances from recombination and collisionally excited lines.
The issue
of temperature fluctuations is an important one, albeit
unresolved. Further details can be found in Peimbert (1995),
Mathis, Torres-Peimbert, \& Peimbert (1998), and Stasi{\'n}ska
(1998).

\section*{Appendix B}

\section*{Abundances From Stellar Spectra}

We do not wish to include a complete description of stellar
atmospheres in this review as most readers have completed a stellar
atmospheres course or have textbooks like Mihalas (1978) on their
shelves or have read shorter introductions like chapter 12 of Cowley
(1995).  However, in the interest of keeping up-to-date, we provide a
brief description with emphasis on oft-heard buzzwords.
Converting measured absorption feature strengths into abundances
requires a model stellar atmosphere that gives the run of physical
variables like temperature and pressure with optical depth. With
today's fast computers, it is no
longer much of a computational burden to compute fairly realistic
LTE model atmospheres, most of which have the following features.
(1) Plane-parallel geometry. This is a good approximation for
most stars, but a spherical geometry is needed for M giants and other
stars that have extended envelopes. (2) A modern model atmosphere 
(e.g. Gustafsson 1989; Kurucz 1993) will
be line-blanketed. That is, individual transitions from (usually
millions) of atomic and molecular lines are explicitly included in the
frequency-dependent opacity calculations along with the various
sources of continuous opacity like electron scattering or H$^-$. (3)
Convection is often treated in the mixing length approximation, but
alternatives are always being tested. (4) For many stars, local
thermodynamic equilibrium (LTE) is assumed. In this case
the local radiation field and the local thermodynamic state (which can
be described by temperature and pressure alone) of the
matter are equilibrated. The alternative is non-LTE (NLTE), in which
the radiation field decouples from the matter. The complicating part
of this is that atomic level occupation numbers then depend mostly on
the radiation field rather than the local matter thermodynamics, and
the radiation field {\em is not local} and therefore a global
self-consistent solution must be sought. Extreme NLTE prevails
in the case of nebulae and hot stars, while most stars can be
treated with the LTE assumption for most lines. However, a temperature
inversion above the photosphere (i.e. a chromosphere) will 
introduce certain NLTE effects in some lines. If these lines are going to
be used for abundance analysis, they need to be treated
appropriately. See Mihalas
(1978) or Kudritzki \& Hummer (1990) for a description of NLTE methods
in hot stars. 

Except for those stars where spherical geometry effects become
important, effective temperature $T_{\rm eff}$, surface gravity log
$g$ (where $g$ is expressed in cm s$^{-2}$), and abundance are
sufficient parameters to begin the calculation of a model
star. Temperatures can be gotten from calibrations of broad-band or
narrow-band colors, by Balmer line strength, or by consideration of two or more ionization states
of the same species (by requiring consistent abundance results from
all ionization states). Abundance analysis is often not very sensitive
to surface gravity.

There is also the matter of Doppler broadening of lines. Thermal
motion of gas particles causes a broadening of the line profile that
is almost always larger than the width of the line as broadened by
pressure and by radiation damping.  In
the LTE approximation inclusion of this effect is trivially
accomplished by convolving the intrinsic Lorentzian line profile with
the Gaussian thermal Doppler profile. But there is also the matter of
bulk motions in the atmosphere, in the sun seen as prominences,
spicules, granules, acoustical waves, and other phenomena. These add
Doppler width to lines, but their velocity distribution is not
known. In the face of the total unknown, we follow historical
precedent and assume Gaussian random motion,
and an empirically adjusted term is added to the width of the Gaussian
Doppler profile. That is, inside the Gaussian broadening function
$e^{-(\Delta\lambda / \Delta\lambda_D)^2}$, 
\begin{equation}
\eqnum{B1}
{{\Delta\lambda_D}\over{\lambda}} = {{1}\over{c}}\cdot
{{2RT}\over{\mu}} + \xi_t^2
\end{equation}
where $\lambda$ is the central wavelength of the line, $\Delta\lambda_D$ is the Doppler width, $R$ is the gas constant, $T$ is the temperature, $\mu$ is the mean atomic weight, $c$ is the light speed, and
$\xi_t$ is called the {\em microturbulent velocity}, meant to
account for Gaussian-random, optically thin turbulent motion in the
atmosphere. The final Doppler motion to be  considered is {\em
macroturbulence}, very large moving elements that can be considered
independent atmospheres. Such motions will not change the equivalent
widths of absorption lines, but they will change the line profile.

Atomic (or molecular) parameters are often crucial to a reliable
abundance. Parameters may include damping constants for radiation and van
der Waals forces. The damped line profile is usually approximated by a
Lorentzian function. The atomic parameters that are mentioned most
often are the ``$gf$ values'', where $g$ is the statistical weight of
the level and $f$ is the oscillator strength; They are
usually combined as ``log $gf$.'' The $gf$ values can be
calculated with varying degrees of accuracy, and many can be measured
in the laboratory, but the usual method of getting accurate values is to
compute a model atmosphere for the sun. The equivalent widths of the
line transitions one is interested in studying are measured from the
solar atlas, and also computed from the model atmosphere using
standard meteoritic + photospheric abundances, for example those of
Grevesse, Noels, \& Sauval (1996). The $gf$ values are then adjusted until the
model equivalent widths match those of the sun. This method begins to
fail for stars too dissimilar from the sun in temperature because the
line of interest will be invisible in one of the two stars, or on a
different part of the curve-of-growth.

The curve-of-growth is the increase in equivalent width of a line as a
function of abundance.
When lines are weak, they grow linearly with the abundance. When the
line center nears its maximum depth, little growth in equivalent
width is seen until the line is so saturated that the weak wings of
the line profile contribute. Thus the textbook curve-of-growth has a
linear-with-abundance portion, an almost flat portion, and then a
proportional-to-the-square-root-of-the-abundance portion. For accurate
abundance work, it is therefore advantageous to choose weak lines in
the linear-growth part of the curve.

Observational material for spectroscopic abundance work will vary
according to circumstance.  In the ideal case, one would have high
S/N, high resolution ($R=\lambda/\Delta\lambda = $ 50000 to 100000)
spectra compared with near-perfect models. In reality, the models have
defects and many/most of the interesting stars are too far away for
both high S/N and high resolution spectra. The usual solution is to go
to lower resolution. The ultimate low-resolution spectrum is
broad-band photometry ($R \sim 10$), which can be coaxed to yield good
information about $Z$ with suitable assumptions and calibrations, but
little information about individual elements. A resolution at which
some elemental abundance information becomes available is about $R
\sim 500$, but only for the strongest, cleanest absorption
features. This is the approximate resolution needed for integrated
starlight studies.

\section*{Appendix C}

\section*{Abundances From Integrated Light} 

If a galaxy is too distant for individual stars to be
spectroscopically analysed one might still hope to learn something
from the properties of the {\it integrated} starlight summed over all
the stars along the line of sight. Broad-band colors or low resolution
spectra can be readily obtained for galaxies of sufficient surface
brightness. (The surface brightness constraint disqualifies dwarf
spheroidals and other low surface brightness galaxies). The question
then becomes how to interpret a radial color profile or a radial
absorption line strength profile in terms of abundance. This is done
by examining the behavior of ``simple'' stellar populations with age
and abundance. A ``simple'' stellar population is an abstraction of a
star cluster, characterized by a single age and a single
abundance. One could use real cluster spectra as templates, if one
knew the clusters' ages and abundances (e.g. Bica 1988) and if the cluster library brackets the age-metallicity space of interest. The
more usual approach is to use a theoretical stellar evolutionary
isochrone as the template population. The isochrone gives direct (but
theoretical) information on age and abundance.

The first point to realize is that a young stellar population will
contain massive stars. O-type stars for the first few tens of millions
years, B-type stars for the first few hundred million years, A-type
stars until roughly 1 billion years of age. These hot stars make the
integrated colors blue, and absorption line strengths weak (except for
hydrogen lines, which are strongest in A-type stars). Furthermore,
young populations are very bright compared to old populations, so they
dominate the integrated light if they are present. In the case of 
spiral galaxies, with their ongoing star formation, getting abundance
information from integrated starlight is
difficult or impossible because the depth of the absorption features
is very strongly modulated by age effects. 

Integrated light is useful, therefore, only in ``dead'' stellar
populations where star formation has not occured for some time, so
that the OBA-type stars are gone. Most of the light (at optical
wavelengths) then comes from FG-type stars at the main sequence
turnoff, and KM-type stars on the red giant branch, in roughly equal
proportions. Practically speaking, this means
that the bulges of spirals and E and S0 galaxies are the targets for
integrated light abundance work. 

Even in old populations, age effects are the major complication in
getting abundance information. One can never be sure that traces of
young stars are completely absent, and if they are present they skew
the abundance result toward lower values because the metal lines will
be weaker. Most isochrone synthesis models (Aaronson et al. 1978;
Worthey 1994; Bressan et al. 1994; Vazdekis et al. 1996)
show that age effects are muted compared to abundance effects in their
impact on the resultant spectrum shape, colors, and line
strengths. The null-spectral-change line is
$\Delta$~log(Age)/$\Delta$~log~$Z \approx -{{3}\over{2}}$, so that a
factor of 3 age change produces the same spectral change as a factor
of 2 change in metallicity. This $-{{3}\over{2}}$ slope is
approximate, and can be different (but not wildly so) for different
colors or spectral indices. In global terms, this is not so bad.
For example, if you
can be reasonably certain that an object formed in the first half of
the universe's history, then your age will be, at most, a factor of
100\% uncertain. This translates to a factor of ${{2}\over{3}}\times
100 = 66$\% uncertainty in abundance, if your isochrone model is
calibrated correctly. This is 0.2 dex --- an impressive accuracy given
the large age range allowed. The qualification ``if your isochrone
model is calibrated correctly'' is important, because scatter among
different models amounts to $\pm 35$\% in age (or roughly $\pm 25$\%
in $Z$ via the ${{3}\over{2}}$ rule). (Charlot et al. 1996)

Models for integrated light are conceptually simple addition problems,
but the ingredients rest on complicated input physics. Stellar
evolutionary isochrones are computed from evolutionary tracks of
different masses in order to construct a snapshot of the population at
a single age. The tracks, in turn, depend on opacities, equations of
state, theories of convection and mass loss, and the numerical methods
of making a model star. With the addition of an initial mass function
the isochrone specifies star
number, luminosity, mass, and temperature in stellar bins along
the curve. The integrated luminosity
at a single wavelength is $ L_{\lambda} = \sum_{\rm isochrone\ bins}
l_{\lambda ,{\rm bin}} dN $, where the number of stars $dN$ is
obtained from an initial mass function $\Phi (M)$ via $dN = \Phi (M)
dM$, and where $l_{\lambda ,{\rm bin}}$ is the monochromatic
luminosity of one star in one bin on the isochrone.  Perhaps most
commonly these days, $l_{\lambda ,{\rm bin}}$ is obtained from a grid
of theoretical model atmosphere fluxes, but empirical libraries are
also used.  

In dealing with integrated light models, the critical parameter is
temperature: for a good model one must make sure that the temperatures
along the isochrone are correct at all points, and that the conversion
from temperature to color (or flux or line strength) is
solid. Relatively small errors in stellar color can propagate almost
unattenuated to the integrated colors (Worthey 1994).

\clearpage

\begin{deluxetable}{lcccccccc}
\tablecolumns{9}
\tablewidth{0pc}
\tablenum{1}
\tablecaption{Milky Way Nebular Studies}
\tablehead{
\colhead{Object} &
\colhead{Author\tablenotemark{1}} &
\colhead{No.} &
\colhead{Range (kpc)} &
\colhead{${O}\over{H}$} &
\colhead{${N}\over{O}$} &
\colhead{${Ne}\over{O}$} &
\colhead{${S}\over{O}$} &
\colhead{${Ar}\over{O}$}
}
\startdata
H II (OPT,RAD)&Shaver&21&5.9-13.7&21&20&9&7&16\\
H II (FIR)&Simpson&22&0.0-10.2&22&22&17&22&\nodata\\
H II (FIR)&Afflerbach&34&0.0-11.4&34&34&\nodata&34&\nodata\\
H II (OPT)&V{\'i}lchez&18&11.7-18.0&9&9&\nodata&6&\nodata\\
H II (OPT)&Fich&18&11.5-17.9&4&4&\nodata&\nodata&1\\
H II (FIR)&Rudolph&5&12.9-17.0&5&2&\nodata&4&\nodata\\
PNe II (OPT) & Maciel&91&4.6-12.4&91&\nodata&76&77&73\\
SNR (OPT)&Fesen&13&4.6-13.5&13&13&\nodata&13&\nodata\\
\enddata
\tablenotetext{1}{Shaver: Shaver et al. (1983); Simpson: Simpson et al. (1995); Afflerbach et al. (1997); V{\'i}lchez: V{\'i}lchez \& Esteban (1996); Fich: Fich \& Silkey (1991); Rudolph: Rudolph et al. (1997); Maciel: Maciel \& K{\"o}ppen (1994); Fesen: Fesen, Blair, \& Kirshner (1985)}  
\end{deluxetable}

\clearpage

\begin{deluxetable}{lccc}
\tablecolumns{4}
\tablewidth{0pc}
\tablenum{2}
\tablecaption{Milky Way Oxygen Gradients}
\tablehead{
\colhead{Author} &
\colhead{G(dex/kpc)} &
\colhead{A$_{8.5}$} &
\colhead{c}
}
\startdata
Shaver&-0.05$\pm$0.01&8.77$\pm$0.14&-0.69\\
Simpson&-0.06$\pm$0.02&8.57$\pm$0.15&-0.56\\
Afflerbach&-0.06$\pm$0.01&8.61$\pm$0.09&-0.66\\
Maciel&-0.07$\pm$0.01&8.66$\pm$0.06&-0.70\\
Fesen&-0.04$\pm$0.03&8.63$\pm$0.32&-0.31\\
Composite&-0.06$\pm$0.01&8.68$\pm$0.05&-0.63
\enddata
\end{deluxetable}

\clearpage

\begin{deluxetable}{lccc}
\tablecolumns{4}
\tablewidth{0pc}
\tablenum{3}
\tablecaption{Heavy Element Abundance Ratio Averages\tablenotemark{1}}
\tablehead{
\colhead{Sample\tablenotemark{2}} &
\colhead{log Ne/O} &
\colhead{log S/O} &
\colhead{log Ar/O}
}
\startdata
Shaver (9,7,16) & -0.69$\pm$0.16 & -1.40$\pm$0.19 & -2.29$\pm$0.16 \\
Maciel (76,77,73) & -0.66$\pm$0.12 & -1.70$\pm$0.22 & -2.27$\pm$0.14 \\
Garnett ( 0,36,0) & \nodata & -1.62$\pm$0.14 & \nodata \\
van Zee (56,173,129) & -0.62$\pm$0.18 & -1.48$\pm$0.26 & -2.24$\pm$0.17 \\
Izotov (54,49,53) & -0.72$\pm$0.05 & -1.55$\pm$0.06 & -2.25$\pm$0.09 \\
Total (195,342,271) & -0.67$\pm$0.14 & -1.55$\pm$0.24 & -2.25$\pm$0.15 \\
Solar\tablenotemark{3} & -0.79 & -1.54 & -2.35 \\
Orion\tablenotemark{4} & -0.75 & -1.46 & -1.84 \\
Helix\tablenotemark{5} & -0.48 & -2.49 & -2.17 
\enddata
\tablenotetext{1}{Arithmetic averages, i.e. log mean antilog}
\tablenotetext{2}{Last name of first author of published sample. Numbers in parentheses indicate sample size for Ne/O, S/O, and Ar/O, respectively}
\tablenotetext{3}{Grevesse et al. 1996}
\tablenotetext{4}{Esteban et al. 1998}
\tablenotetext{5}{Henry, Kwitter, \& Dufour 1999}
\end{deluxetable}

\clearpage

\begin{deluxetable}{cccccc}
\tablecolumns{6}
\tablewidth{0pc}
\tablenum{A1}
\tablecaption{Prominent Emission Lines In Nebulae}
\tablehead{
\colhead{Ion} &
\colhead{Wavelength ({\AA})} &
\colhead{Excitation\tablenotemark{1}} &
\colhead{Ion} &
\colhead{Wavelength ({\AA})} &
\colhead{Excitation} 
}
\startdata
C IV & 1549 & C& H I & 4861 & R\\
He II & 1640 & R & [O III] & 4959,5007 & C\\
C III] & 1909 & C & [N II] & 5199 & C\\
\[[O II] & 3727 & C & He I & 5876 & R\\
\[[Ne III] & 3869,3968 & C & [O I] & 6300,6360 & C\\
He I & 3889 & R & [S III] & 6312 & C\\
\[[S II] & 4072 & C & [N II] & 6548, 6584 & C\\
H I & 4101 & R & H I & 6563 & R\\
H I & 4340 & R & [S II] & 6716,6731 & C\\
\[[O III] & 4363 & C & [Ar III] & 7135 & C\\
He I & 4471 & R & [O II] & 7325 & C\\
He II & 4686 & R & [S III] & 9069,9532 & C
\enddata
\tablenotetext{1}{Excitation mechanism, where C=collisional, R=recombination}
\end{deluxetable}

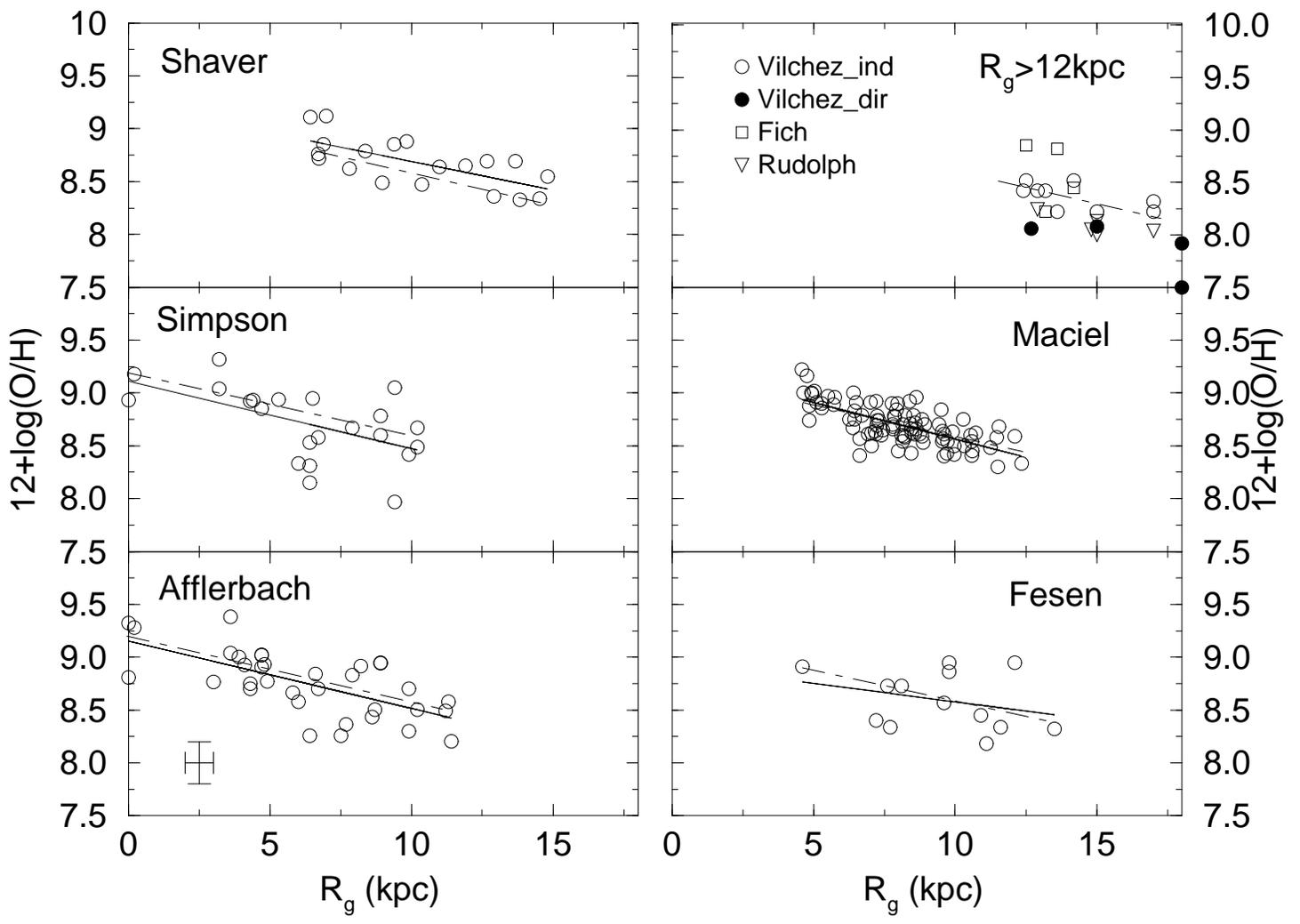
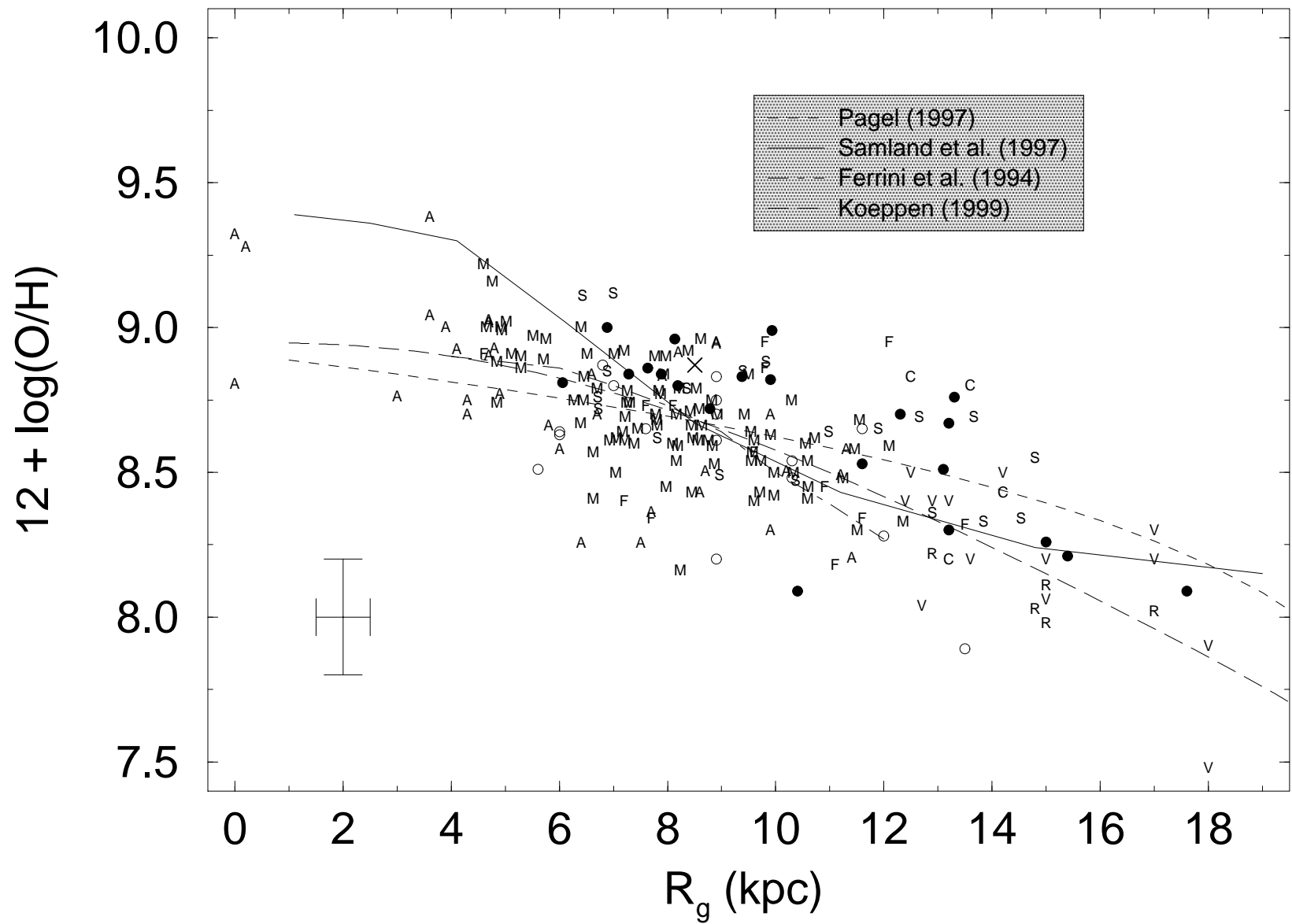
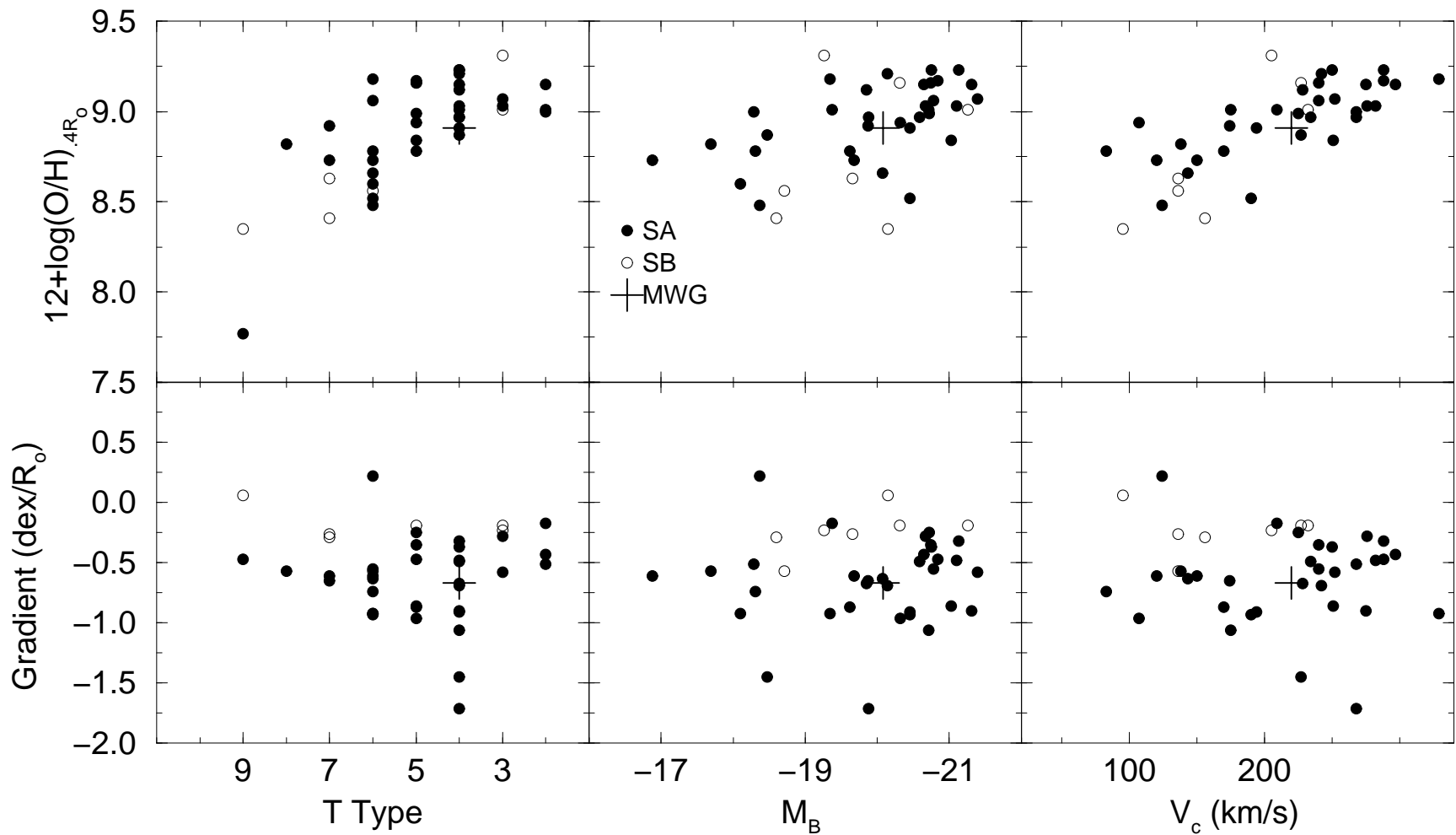
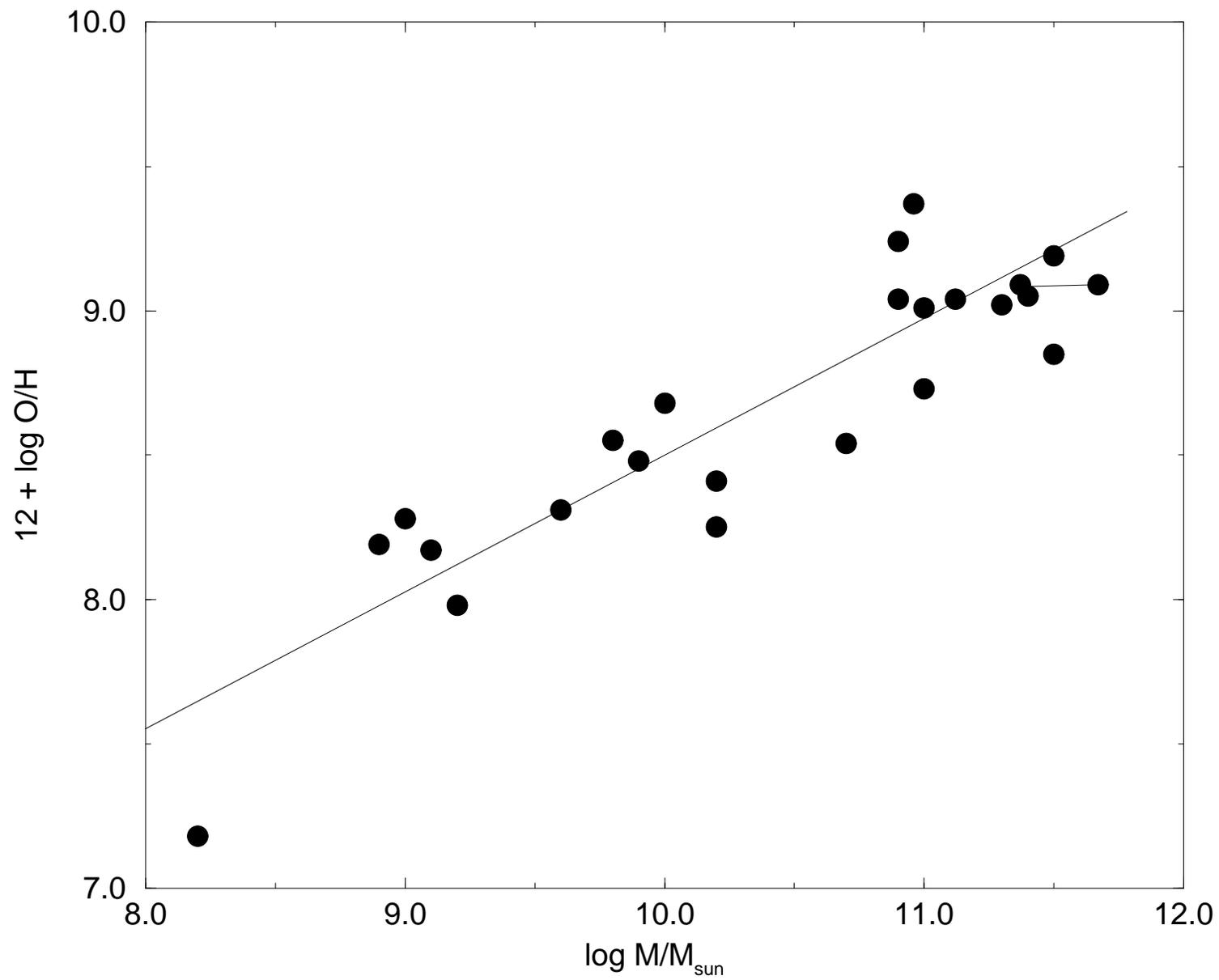
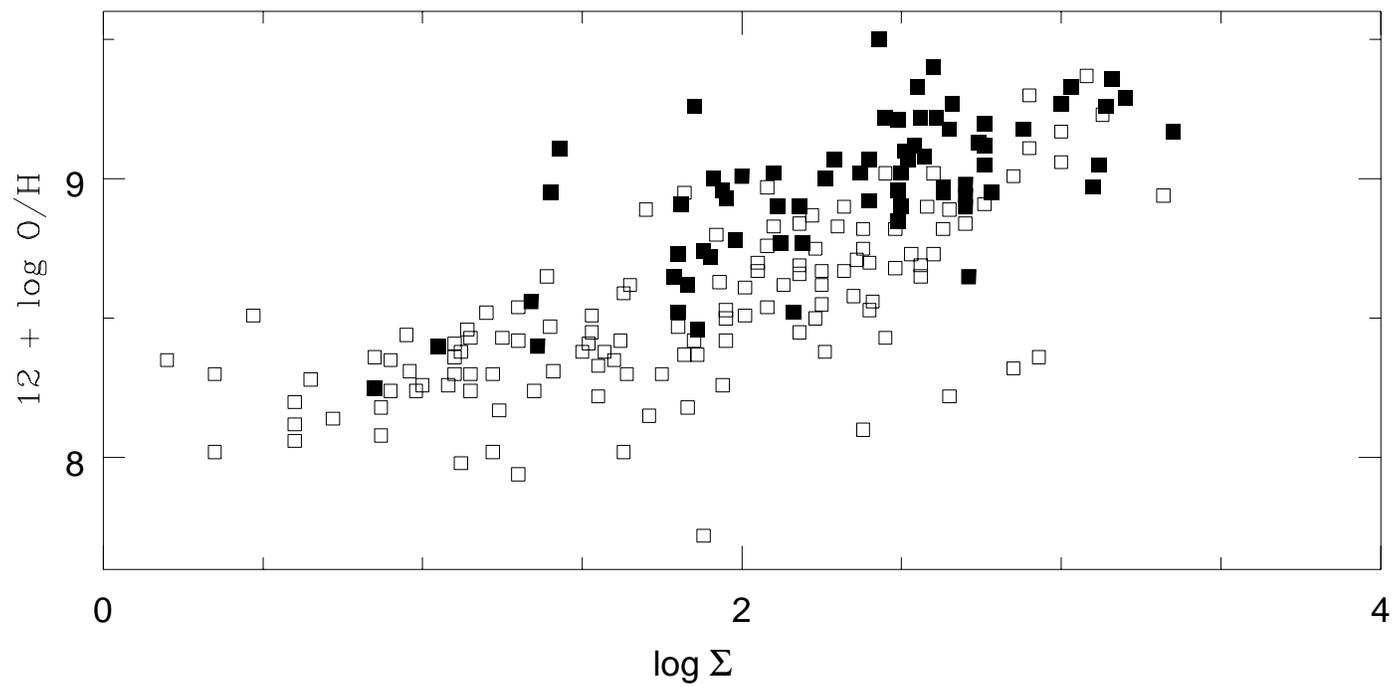
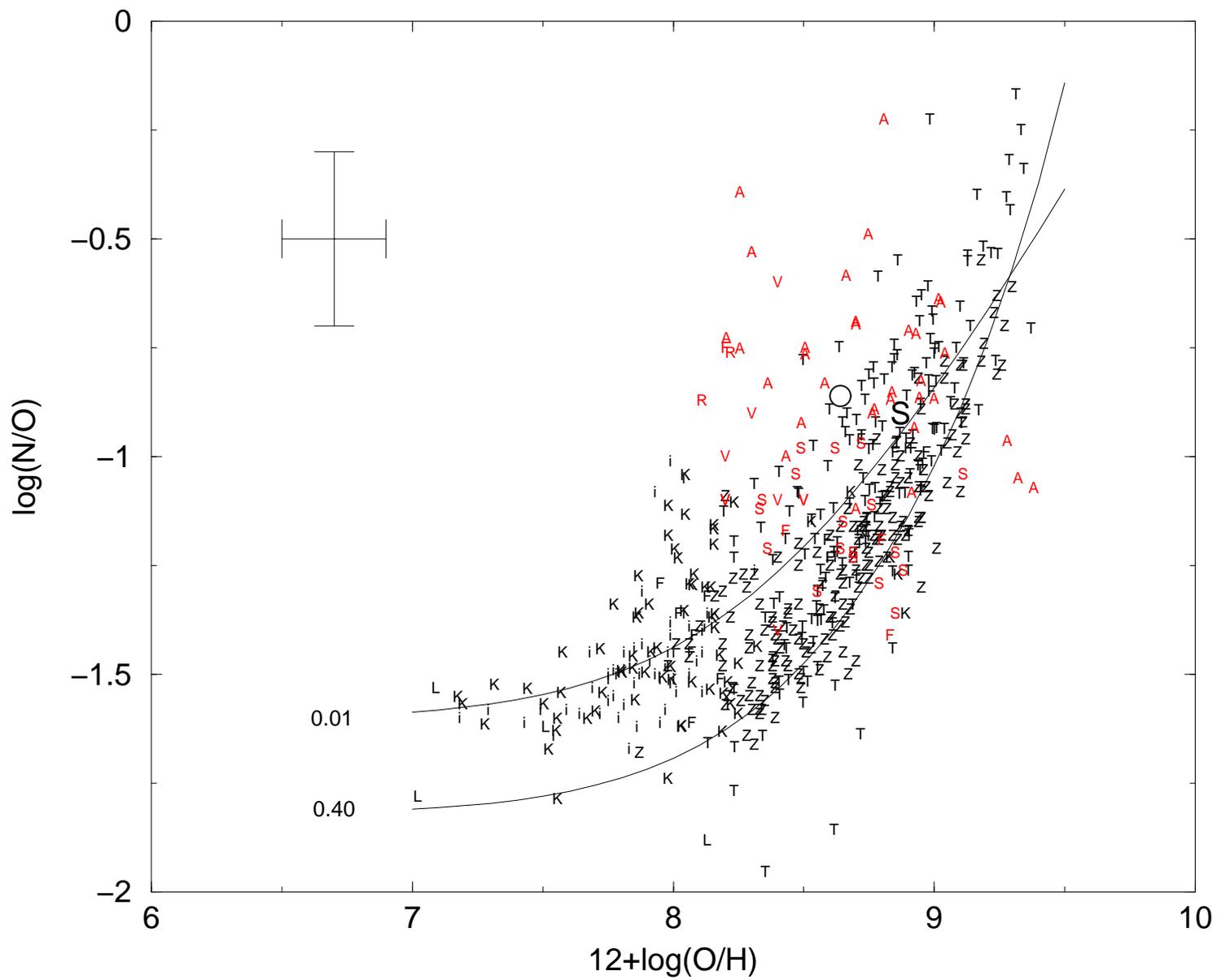
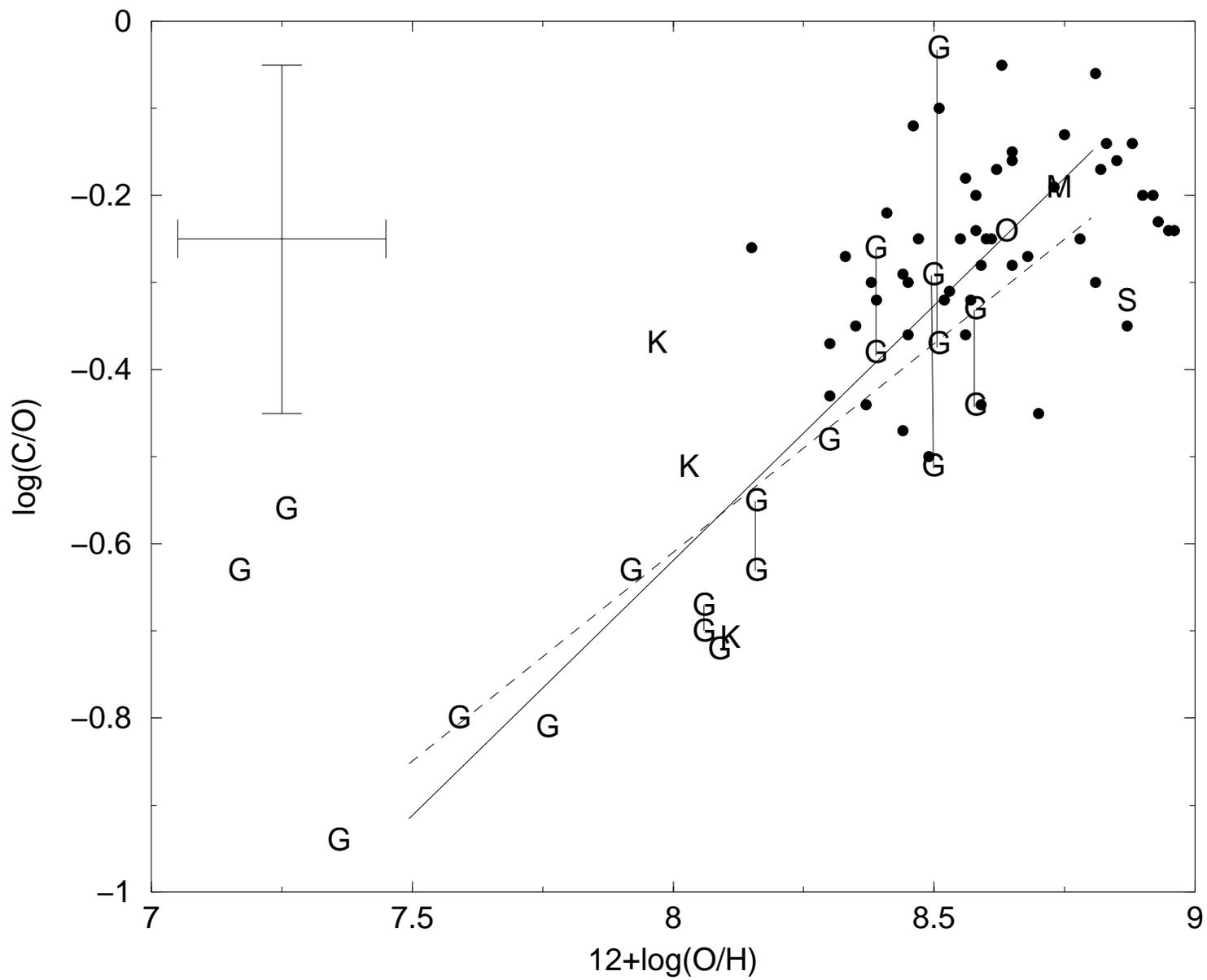
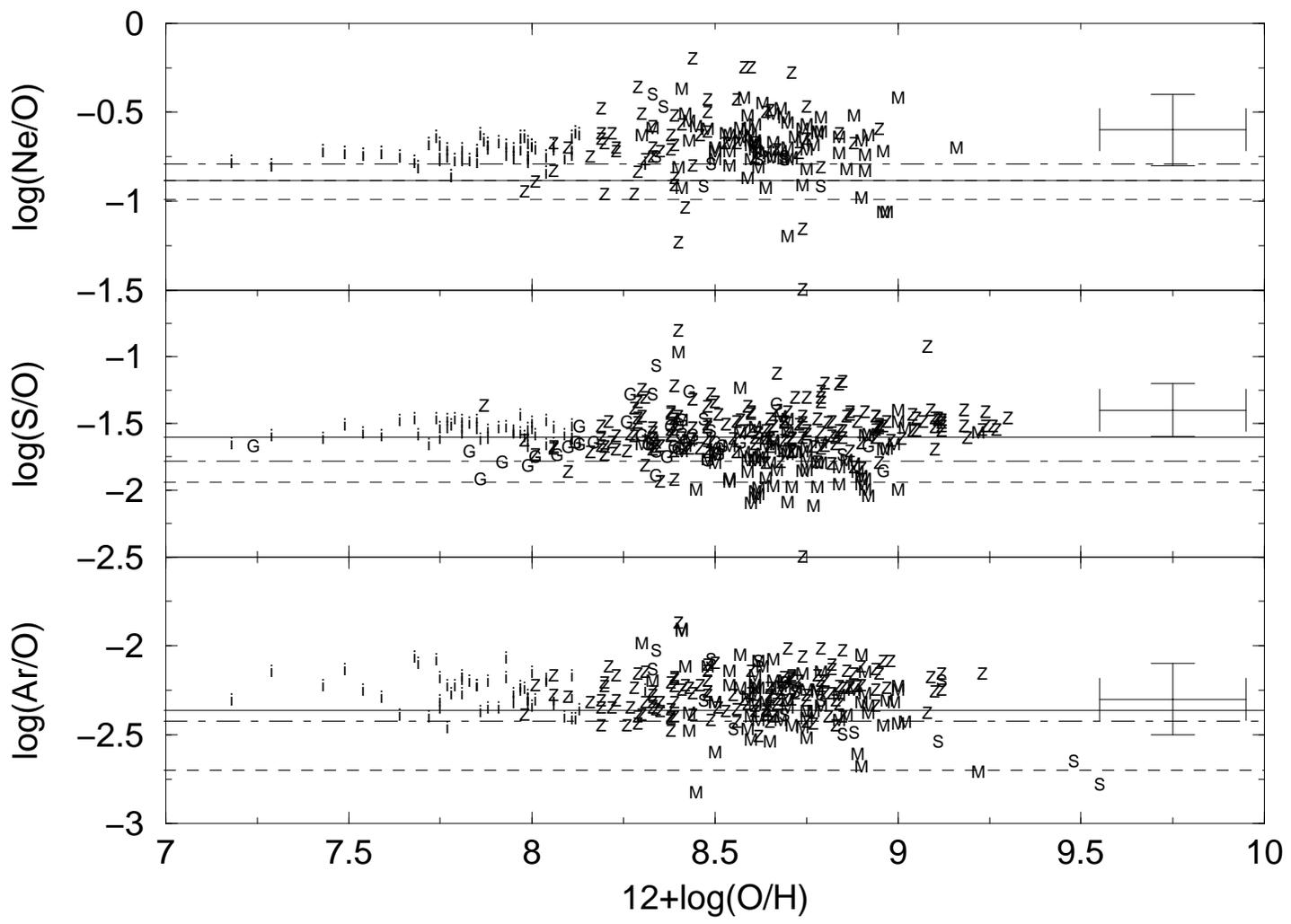
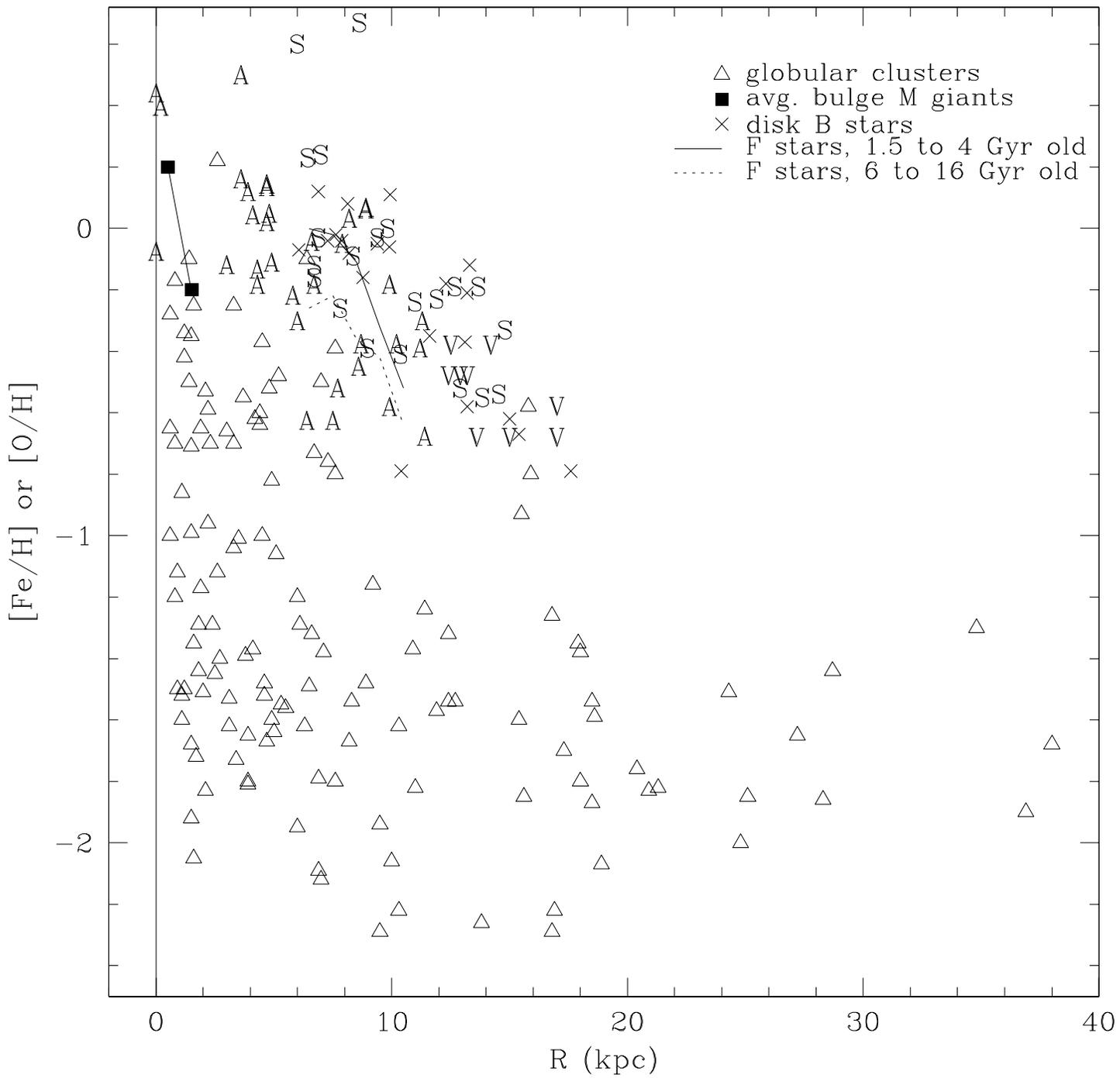
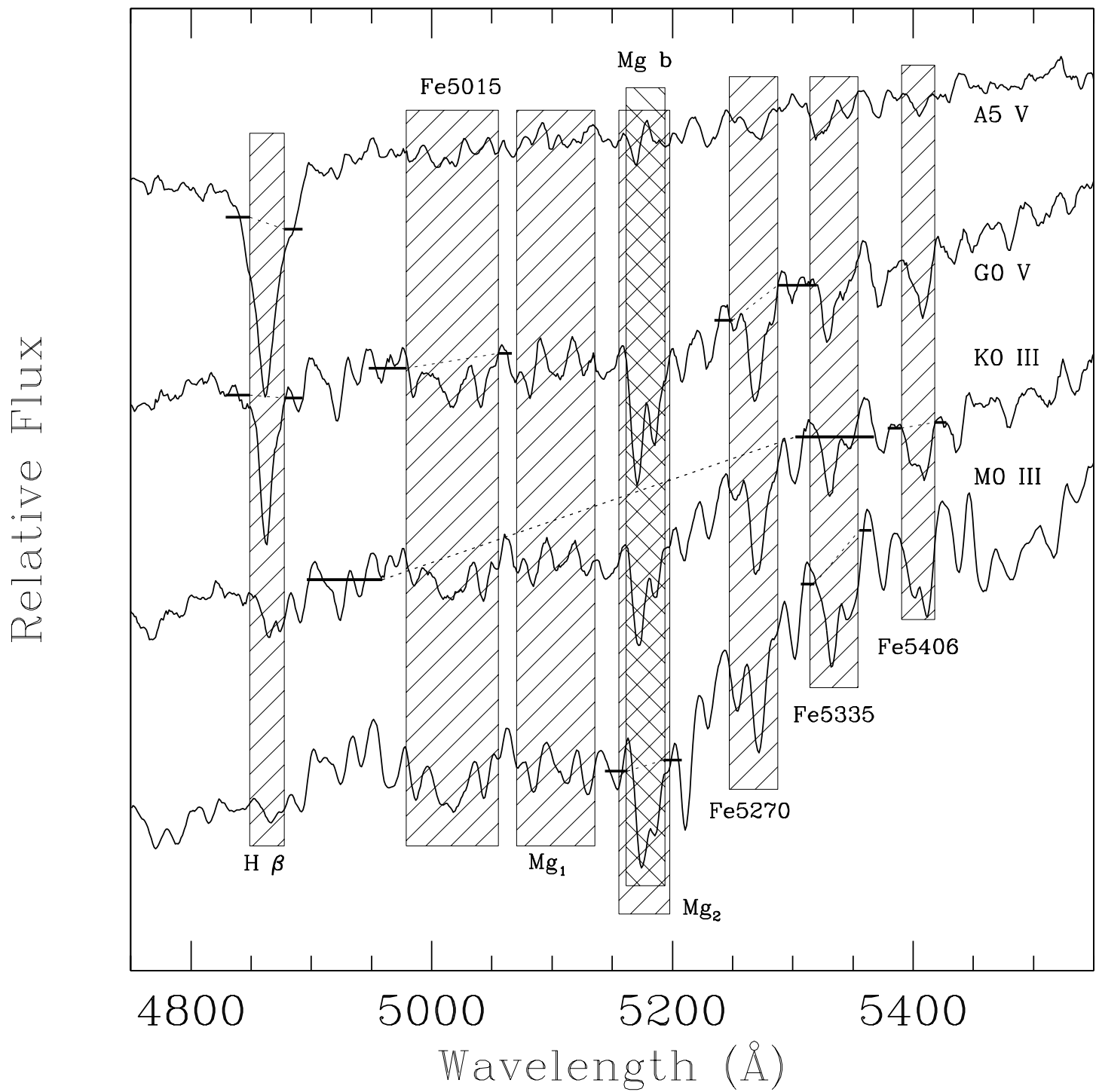
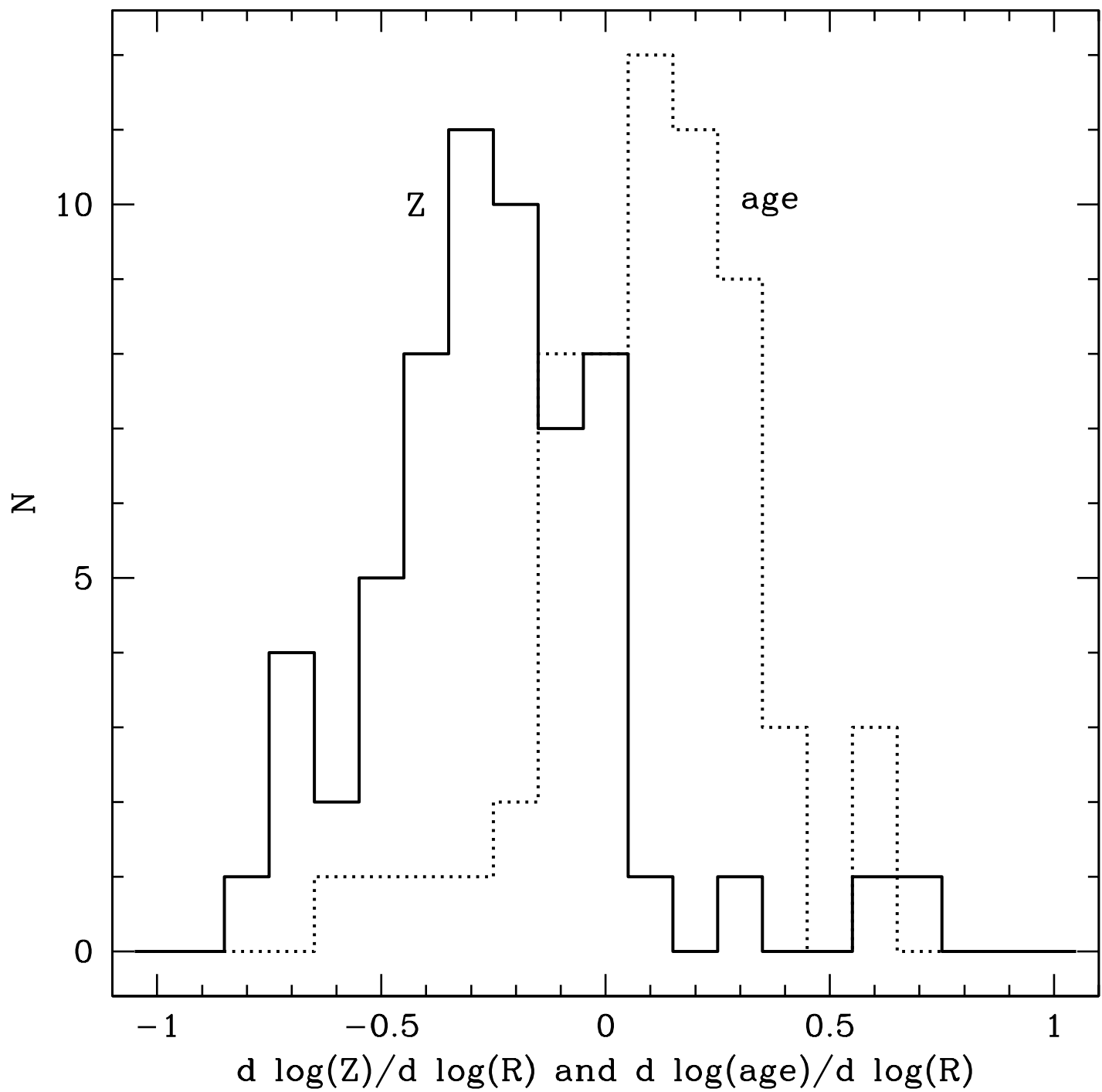
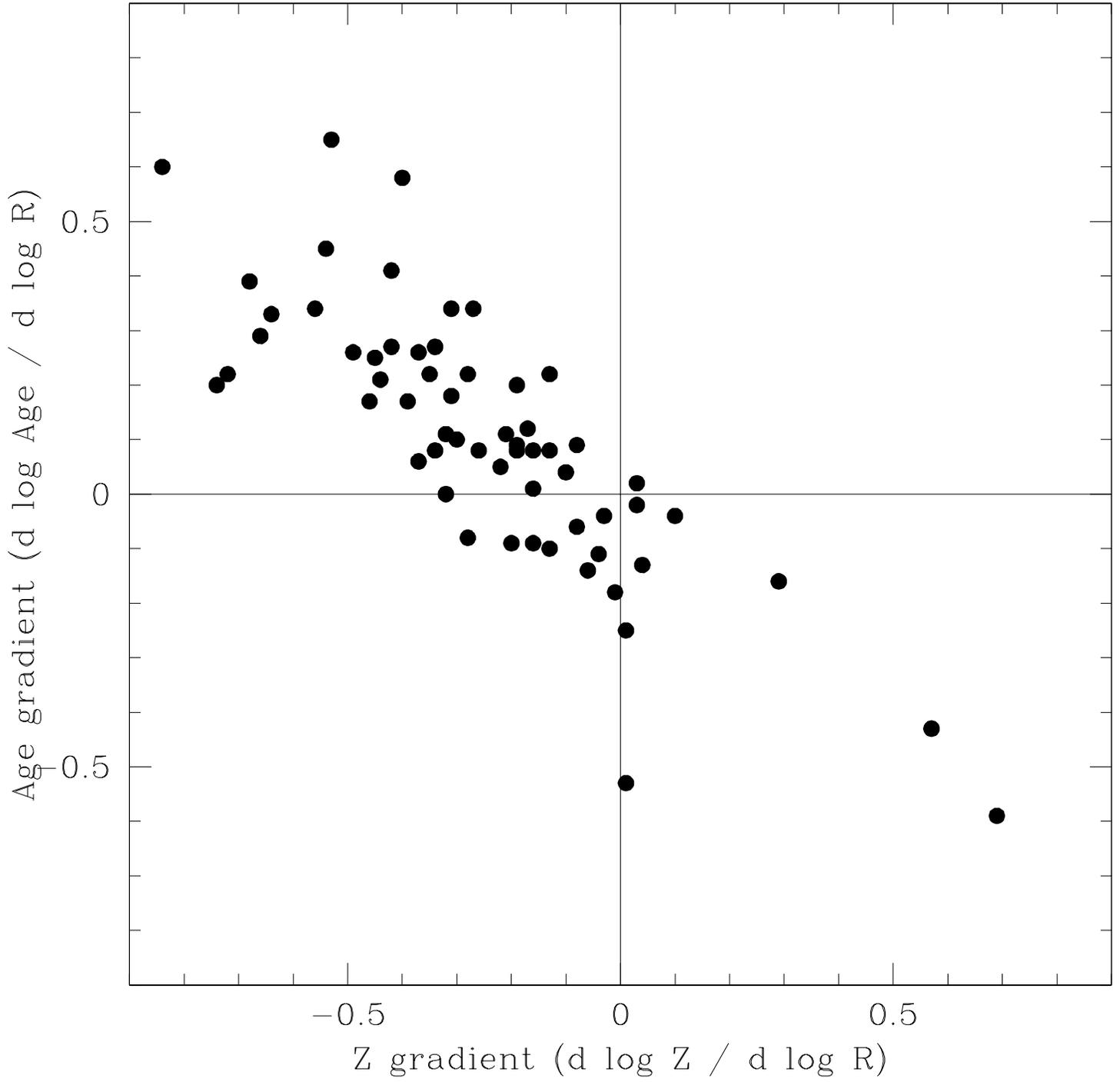
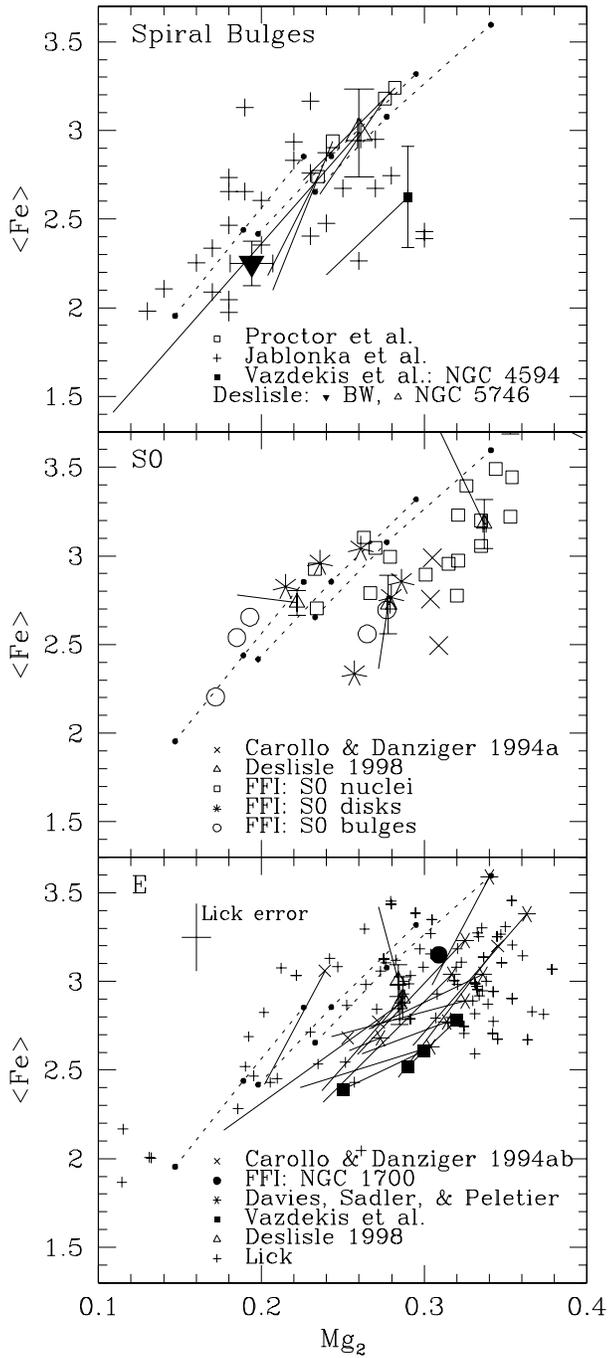
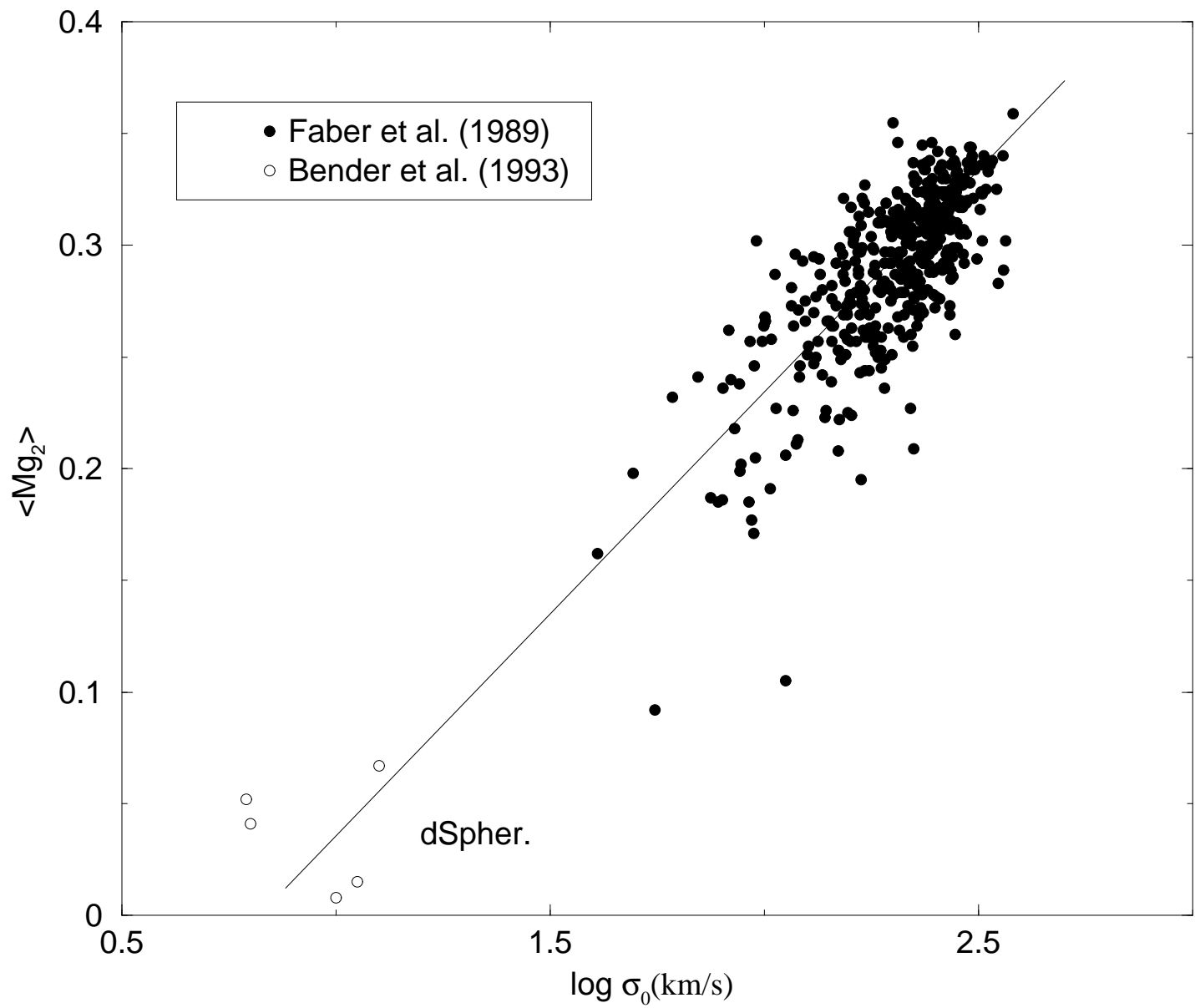
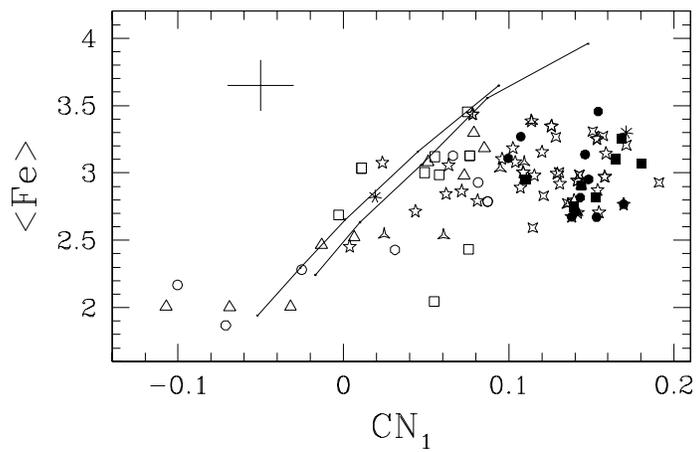
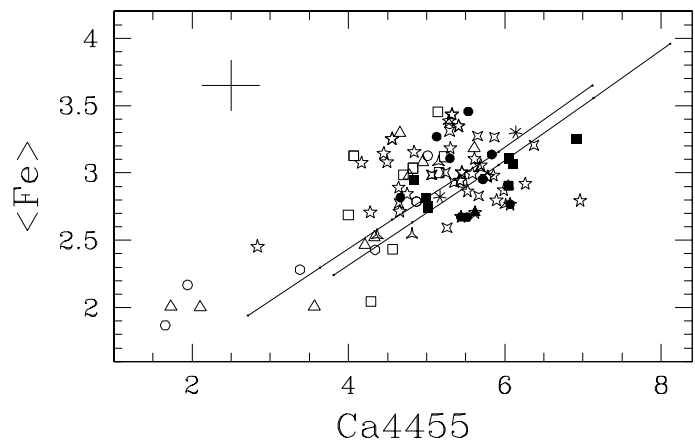
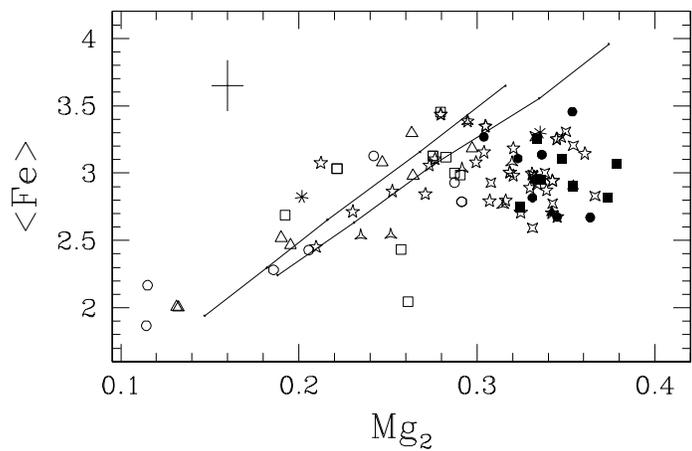
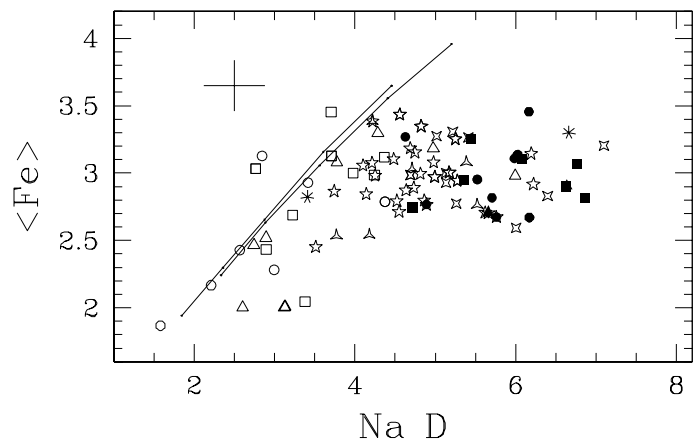
\end{document}